\documentclass[twocolumn]{aastex62}

\usepackage{subfigure}
\usepackage{transparent}
\usepackage{graphicx}
\usepackage{chngcntr}
\usepackage{multirow}
\usepackage{subscript} 
\usepackage{ascii}

\newcommand{\hst}{\textit{HST} }
\graphicspath{{./}{figures/}}

\accepted{for publication in the Astrophysical Journal: \today}

\shorttitle{Leo P}
\shortauthors{Goldman et al.}

\begin{document}

\title{\textbf{Asymptotic Giant Branch Stars in the Nearby Dwarf Galaxy Leo P}\footnote{Based on observations made with the NASA/ESA Hubble Space Telescope, obtained at the Space Telescope Science Institute, which is operated by the Association of Universities for Research in Astronomy, Inc., under NASA contract NAS 5-26555. These observations are associated with program HST-GO-14845.}}

\correspondingauthor{Steven Goldman}
\email{sgoldman@stsci.edu}

\author[0000-0002-8937-3844]{S. R. Goldman}
\affil{Space Telescope Science Institute, 3700 San Martin Drive, Baltimore, MD 21218, USA}

\author[0000-0003-4850-9589]{M. L. Boyer}
\affil{Space Telescope Science Institute, 3700 San Martin Drive, Baltimore, MD 21218, USA}

\author[0000-0001-5538-2614]{K. B. W. McQuinn}
\affil{University of Texas at Austin, McDonald Observatory, 2515 Speedway, Stop C1402, Austin, Texas 78712 USA}
\affil{Rutgers University, Department of Physics and Astronomy, 136 Frelinghuysen Road, Piscataway, NJ 08854, USA}

\author[0000-0003-4520-1044]{G. C. Sloan}
\affil{Space Telescope Science Institute, 3700 San Martin Drive, Baltimore, MD 21218, USA}
\affil{Department of Physics and Astronomy, University of North Carolina Chapel Hill, Chapel Hill, NC 27599-3255}

\author[0000-0003-0356-0655]{I. McDonald}
\affil{Jodrell Bank Centre for Astrophysics, Alan Turing Building, University of Manchester, M13 9PL, UK}

\author[0000-0002-1272-3017]{J. Th. van Loon}
\affil{Lennard-Jones Laboratories, Keele University, ST5 5BG, UK}

\author[0000-0002-3171-5469]{A. A. Zijlstra }
\affil{Jodrell Bank Centre for Astrophysics, Alan Turing Building, University of Manchester, M13 9PL, UK}

\author[0000-0002-2954-8622]{A. S. Hirschauer}
\affil{Space Telescope Science Institute, 3700 San Martin Drive, Baltimore, MD 21218, USA}

\author[0000-0003-0605-8732]{E. D. Skillman}
\affil{Minnesota Institute for Astrophysics, School of Physics and Astronomy, 116 Church Street S. E.,\\ University of Minnesota, Minneapolis, MN 55455, USA}

\author[0000-0002-2996-305X]{S. Srinivasan} 
\affil{Instituto de Radioastronom\'ia y Astrof\'isica, Universidad Nacional Aut\'onoma de M\'exico Antigua Carretera a P\'atzcuaro \\ \#8701 Ex-Hda. San Jos\'e de la Huerta Morelia, Michoac\'an. M\'exico. C.P.58089
}

\begin{abstract}
We have conducted a highly sensitive census of the evolved-star population in the metal-poor dwarf galaxy Leo P and detected four asymptotic giant branch (AGB) star candidates. Leo P is one of the best examples of a nearby analog of high-redshift galaxies because of its primitive metal content (2\% of the solar value), proximity, and isolated nature, ensuring a less complicated history. Using medium-band optical photometry from the \textit{Hubble Space Telescope} (\textit{HST}), we have classified the AGB candidates by their chemical type. We have identified one oxygen-rich source which appears to be dusty in both the \textit{HST} and \textit{Spitzer} observations. Its brightness, however, suggests it may be a planetary nebula or post-AGB object. We have also identified three carbon-rich candidates, one of which may be dusty. Follow-up observations are needed to confirm the nature of these sources and to study the composition of any dust that they produce. If dust is confirmed, these stars would likely be among the most metal-poor examples of dust-producing stars known and will provide valuable insight into our understanding of dust formation at high redshift. 
\end{abstract} 


\section{INTRODUCTION}
\label{sec:intro}

In the advanced stages of their evolution, low- to intermediate-mass stars (0.8 $\lesssim$ \textit{M} $\lesssim$ 8\,M$_{\odot}$) become important producers of dust. Once in the thermally-pulsing phase of the asymptotic giant branch (TP-AGB), these stars can vary dramatically in brightness on timescales between 60 and 2000 days as a result of pulsations, which are a critical component of how they produce dust. Their variable nature allows circumstellar gas to be levitated to large radii near the gravitational barrier, where it cools and condenses into dust and is then driven away by radiation pressure \citep[e.g.][and references therein]{Hoefner2018}. As they produce dust, TP-AGB stars become obscured in the optical and bright in the infrared (IR), as stellar light is scattered and absorbed by the dust and re-radiated in the IR.

AGB stars can have surface chemistries rich in either oxygen- (M-type) or carbon (C-type), which toward their later stages of evolution result in the production of either oxygen-rich or carbon-rich dust grains. AGB stars that are lower mass or are early in their evolution are oxygen-rich stars. As they evolve, their surface chemistry is dictated by the initial mass and metallicity, which affect the efficiency and timescales of internal processes. Third dredge-up (TDU) events cause AGB stars above $\sim$\,1.3--2\,M$_{\odot}$ to become carbon stars, while even more massive AGB stars ($\sim$\,3--5\,M$_{\odot}$) will become carbon stars and then transition back to oxygen-rich stars as a result of hot-bottom burning \citep[HBB;][]{Boothroyd1993}. The mass limit ranges are a result of these internal processes.  The upper and lower mass-limits of carbon stars decrease in more metal-poor samples \citep{Marigo2017}.

From the ground, we can observe molecules like TiO and VO on the surfaces of these stars. From space, however, we can detect additional molecules in the IR  with much more sensitivity, like water in oxygen-rich AGB stars and C$_2$+H$_2$ and HCN in carbon-rich AGB stars \citep{Aoki1999,Cernicharo1999,Volk2000,Kraemer2002}, which allow for categorization of chemical type. We can use them to gain insight into the efficiency of the dust production rate of TP-AGB stars, their dust properties and composition, stellar evolution, and reveal distinct chemical populations.

\subsection{Dust production and metallicity}

Dust reservoirs have been observed in galaxies and quasars out to redshifts $z\,\sim\,6$ \citep{Bertoldi2003,Robson2004,Beelen2006}. The origin of this dust, however, remains unclear. Massive oxygen-rich TP-AGB stars are expected to be able to produce dust 30 Myr after forming (for a 10\,M$_{\odot}$ star), compared to at least 280--380 Myr for carbon stars \citep{Ventura2002,Herwig2004,Volk2000}. The dust production of massive oxygen-rich TP-AGB stars should be affected by metallicity as these stars should not be able to create the raw materials needed to seed the nucleation of dust grains, as opposed to carbon stars \citep{Lagadec2008,Nanni2018}. As a result, the earliest and highest-redshift dust producers (the massive oxygen-rich TP-AGB stars) are expected to be limited by their initial metallicity. 

Observations of nearby carbon-rich TP-AGB samples conflict on the degree that metallicity affects dust production \citep{Loon2006,Loon2008,McDonald2011,Sloan2012,Sloan2016}. For oxygen-rich AGB stars, studies have shown that dust production \citep{Sloan2008,Sloan2010} and expansion velocity \citep{Marshall2004,Goldman2017} are limited in metal-poor AGB stars in nearby galaxies and globular clusters. This, however, is only the case for AGB stars. Supernovae (SNe) may dominate dust production at high redshift and in metal-poor galaxies, yet the level of dust that they destroy (both newly formed and pre-existing) remains unclear \citep{Temim2015,Lakicevic2015}.

\begin{figure*}{}
 \begin{flushleft}
 \includegraphics[width=0.95\linewidth]{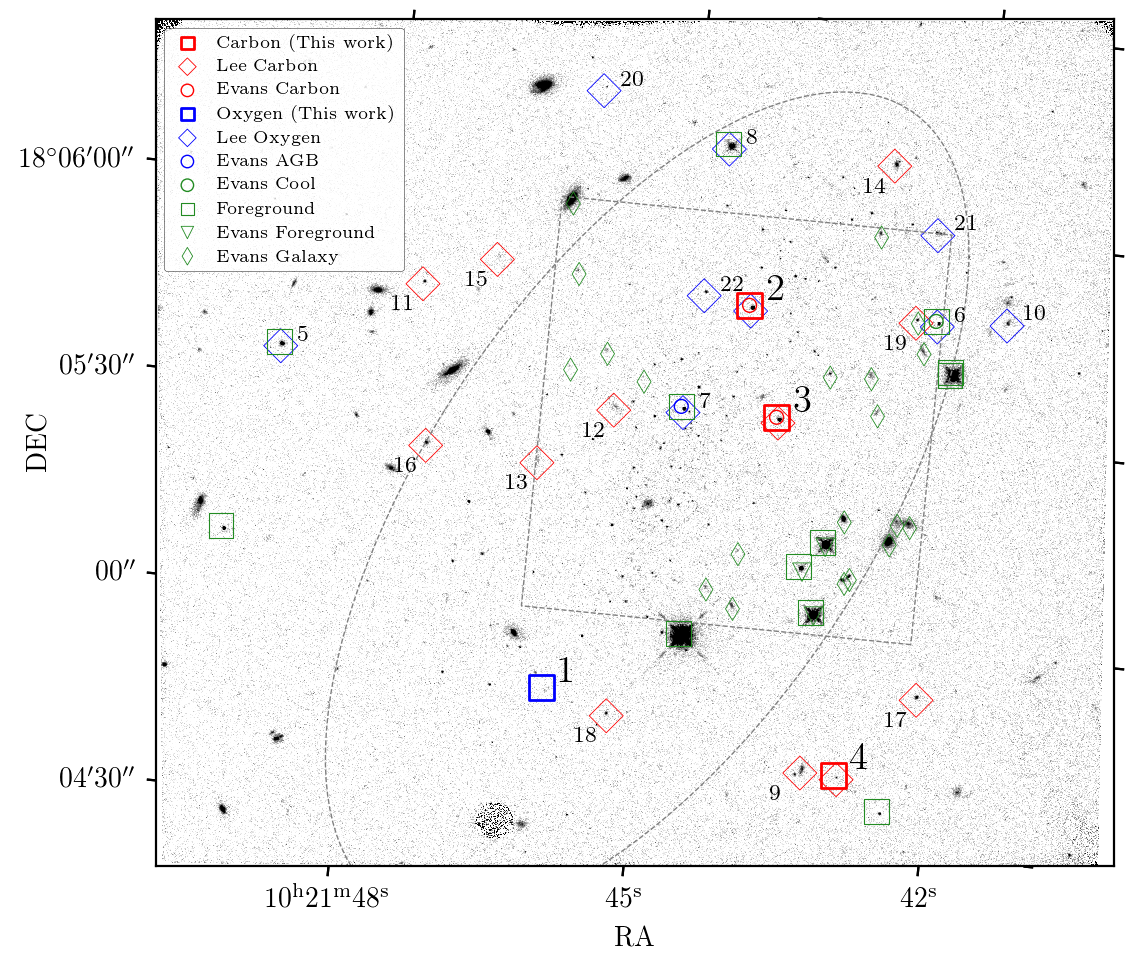}\hspace{0cm}
 \end{flushleft} \vspace{-0.6cm}
 \caption{An \hst F127M medium-band drizzled mosaic of Leo P (half-light radius shown with the gray dashed oval) with sources categorized as oxygen- (blue) and carbon-rich (red). Our AGB classifications as well as those made by \citet{Lee2016} and \citet{Evans2019} are shown with different shapes, where the label ``Evans AGB'' corresponds to the source with a confirmed membership to Leo P using Ca\,\textsc{ii} absorption (Source 22), and ``Evans Cool'' (Source 7) corresponds to the source that was only confirmed as a cool star from Ca\,\textsc{ii} absorption. Green symbols show sources that have been categorized as foreground stars or background galaxies in this work and \citet{Evans2019}. The footprint of the previous VLT MUSE observation by \citet{Evans2019} is shown as the gray dashed square and the Gemini North NIRI observations cover the whole field of view as well as the \textit{Spitzer} and \textit{HST} observations. The oval artifact in the bottom of the image is a known $HST$ artifact. \\}
 \label{hst_image}
\end{figure*}

\subsection{Metal-poor AGB samples}

The largest available samples of evolved stars more metal poor than the Milky Way lie in the Large and Small Magellanic Clouds (LMC, SMC), at around one-half and one-fifth solar metallicity, respectively. Smaller samples have been detected that span a range of metallicities in nearby dwarf galaxies and globular clusters. We require larger samples and in even more metal-poor environments to study the effect of metallicity on the evolution of AGB stars and their contribution of dust to the ISM. 

\paragraph{Dwarf galaxies}
Metal-poor AGB populations have been detected in nearby galaxies spanning 2 dex in metallicity \citep{Battinelli2007,Menzies2008,Whitelock2009,Battinelli2008,Menzies2010,Menzies2011,Lorenz2011,Whitelock2012,McDonald2012,Sloan2012,Bellazzini2014,Menzies2015,Boyer2015b,Boyer2017,Jones2018}. One of these sources is in the Sagittarius Dwarf Irregular Galaxy (Sag DIG), with a gas-phase metallicity of 12\,$+$\,log(O/H)\,=\,7.26\,--\,7.50 \citep{Saviane2002}, and was found to have a pulsation period of 950 days \citep{Whitelock2018}, indicative of a very late stage of evolution and a high progenitor mass, and was additionally found to be oxygen-rich and dusty \citep{Boyer2017}. The DUST in Nearby Galaxies with \textit{Spitzer} \citep[DUSTiNGS;][]{Boyer2015a} and \textit{SPitzer} InfraRed Intensive Transients Survey \citep[SPIRITS;][]{Kasliwal2017} surveys were designed to look for dusty metal-poor long-period variables (LPVs) in the IR and have successfully uncovered populations in nearby galaxies \citep{Boyer2015a,Boyer2015b,Goldman2019a,Karambelkar2019}. Additional observations with the \textit{Hubble Space Telescope} (\textit{HST}) have chemically classified a significant fraction of these \citep{Boyer2017}. These observations have revealed candidate AGB stars in seven galaxies reaching down to metallicities of [Fe/H]\,=\,$-1.85$.

\paragraph{Globular clusters}

Variables have also been detected in globular clusters within the Galaxy \citep{Whitelock1986,Clement2001,Origlia2002,Feast2002,Lebzelter2005,Boyer2006,Matsunaga2007,McDonald2009,McDonald2011b,McDonald2011}, and SMC \citep{Feast2002}. Metal-poor dust-producing AGB stars have also been found in $\omega$ Cen \citep{McDonald2011a}, a cluster with an average metallicity of [Fe/H] $\sim$ $-$1.6 dex \citep{Smith2000}. These are low mass ($<$\,1\,M$_{\odot}$) AGB stars, however, and not representative of the high-mass TP-AGB stars that would contribute dust in high-redshift galaxies.

\begin{deluxetable*}{ccccccccc}
\tabletypesize{\normalsize}
\tablecolumns{9}
\tablecaption{Leo P Observations \label{table:observations}}

\tablehead{
\multirow{2}{*}{Instrument}&
\colhead{Program}& 
\colhead{Filter}&
\colhead{$t$\textsubscript{exp}} & 
\colhead{RA}& 
\colhead{Dec}& 
\colhead{Start Date}&
\colhead{Orient.} &
\multirow{2}{*}{FOV}\\
&
ID &
&
(s) &
(J2000)&
(J2000)&
(UT)&
(E of N) &
}

\startdata
\multirow{2}{*}{\textit{Spitzer:IRAC}} & 
\multirow{2}{*}{13009} & 
[3.6] &
\multirow{2}{*}{400} & 
\multirow{2}{*}{10:21:44.71} & 
\multirow{2}{*}{+18:06:20.3} & 
\multirow{2}{*}{2017 Feb 28 18:17:09} & 
\multirow{2}{*}{\llap{$-$6}6.\rlap{5$^{\circ}$}} &
\multirow{3}{*}{$2 \times (5.2^{\prime} \times 5.2^{\prime})$} \\ 
\vspace{1mm}
 & & [4.5] & & & & &  \\
\multirow{3}{*}{\textit{Hubble:WFC3/IR}} & 
\multirow{3}{*}{14845} & 
F127M & 
797 &
\multirow{3}{*}{10:21:45.32} & 
\multirow{3}{*}{+18:05:26.7} & 
\multirow{3}{*}{2017 Mar 10 07:54:39} & 
\multirow{3}{*}{5.\rlap{7$^{\circ}$}} &
 \\
 & & F139M & 847 & & & & & $2.27^{\prime} \times 2.27^{\prime}$ \\
 & & F153M & 797 & & & & 
\enddata 

\tablenotetext{}{\small{\textbf{Note.} The field of view (FOV) of the \textit{IRAC} detector has two chips separated by 1.52$^{\prime}$ }} 
\end{deluxetable*} 

\subsection{Leo P}
\label{sec:leo_P} 

To search for high-redshift AGB analogs we have observed the nearby dwarf galaxy Leo P (shown in Figure \ref{hst_image}), which is the most metal-poor gas-rich galaxy resolvable with current instruments. Leo P is a dwarf galaxy with a stellar mass of $5.7 \times 10^{5}$\,M$_{\odot}$ and distance of 1.62\,$\pm$\,0.15 Mpc \citep{McQuinn2013,McQuinn2015}. Observations have shown that the ratio of gas to stars is 2:1 and the ratio of total mass to baryonic mass is 15:1 \citep{Bernstein-Cooper2014}. Originally discovered by the Arecibo Legacy Fast ALFA survey \citep[ALFALFA;][]{Giovanelli2005}, observations have revealed the galaxy's stellar populations \citep{McQuinn2015} as well as a luminous star-forming H\,\textsc{ii} region \citep{Rhode2013}. One of the most remarkable aspects of Leo P, however, is its ISM metallicity \citep[12+log(O/H)\,=\,7.17\,$\pm$\,0.04;][]{Skillman2013}, which is essentially identical to that of the well known low metallicity galaxy I Zw 18 \citep[e.g.][]{Skillman1993}, but is ten times closer. The low metallicity is also evident in the stellar population at 1.8\% solar \citep[{[Fe/H] $=$ $-$1.8 $\pm$ 0.1 dex; }][]{McQuinn2015}.

Leo P has shown a nearly constant star formation rate (SFR) over the last 4 Gyr. Over the last 200 Myr the SFR of Leo P has been $\sim$\,$6.5 \times 10^{5}$ M$_{\odot}$\,yr$^{-1}$ \citep{McQuinn2015}, making it the most metal-poor star-forming galaxy in the Local Volume. Searching for AGB stars in star-forming galaxies is critical, as the short evolutionary timescales of the massive dust-producing AGB stars are limited to galaxies with recent star formation. Using the Color-Magnitude Diagram (CMD)-fitting tool {\asciifamily MATCH} \citep{Dolphin2002}, we have fit a synthetic CMD of Leo P and calculated the expected number of massive AGB stars ($M > 5$\,
M$_{\odot}$) to be around three. This code has the added benefit of randomly sampling the Initial Mass Function (IMF), which is important for galaxies with low SFRs like Leo P.  

Leo P has shown to lie near to the luminosity-metallicity and stellar mass-metallicity relationships for dwarf galaxies. Unlike all of the other extremely metal-poor (XMP) galaxies like I\,Zw\,18 \citep{Berg2012}, however, Leo P is not offset from the luminosity-metallicity relation \citep{Skillman2013}. This suggests that Leo P's star-formation history (SFH) is more similar to a high-redshift galaxy than any other stellar sample resolvable with current instruments. Leo P is also the only known XMP dwarf galaxy that lacks a complicated history, which allows us to better constrain the different processes that can affect the abundances of metals in dwarf galaxies. Based on the age--metallicity relationship of \citet{Rafelski2012}, Leo P has abundances representative of a galaxy with a redshift of $z$\,$\gtrsim$\,3.2, or a galaxy in an epoch 11.7 Gyr ago. While Leo P differs from high-redshift galaxies in its lower-mass evolved star population and enrichment history, it is an excellent environment to study the effects of metallicity on stellar evolution and dust production.  

\begin{figure}
 \hspace{-0.5cm}
 \includegraphics[width=\linewidth]{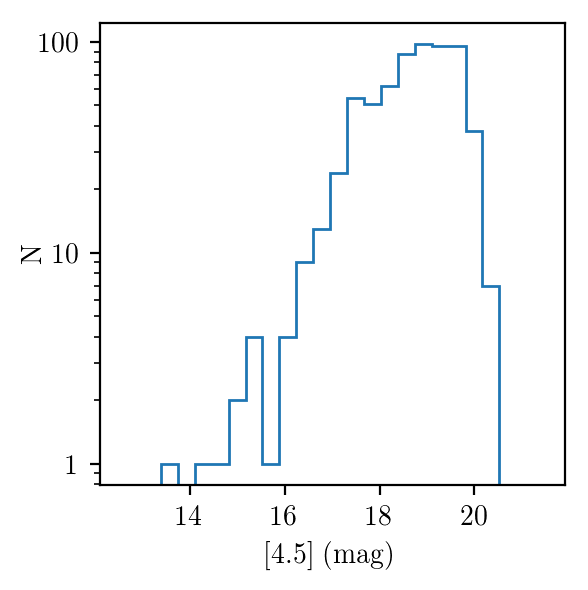}
 \caption{The luminosity function of the \textit{Spitzer} [4.5] photometry. The histogram shows a drop in sensitivity below $\sim18$\,mag.} 
 \label{fig: luminosity function} 
\end{figure}

In this paper we present the results from a highly sensitive census of the evolved star population in the metal-poor dwarf galaxy Leo P. We have detected four AGB candidates, two of which show evidence of dust. One of these dusty sources may be a planetary nebula (PN) or post-AGB star that has been producing oxygen-rich dust. The other dusty source is likely a dusty carbon star. Section 2 describes the observations and methods used to identify the AGB candidates. Section 3 presents the AGB candidates, compares them with previous observations, and discusses how Leo P compares to a high-redshift galaxy.

\begin{deluxetable*}{ccccccccccccccc}
\tablewidth{\linewidth}
\tabletypesize{\small}
\tablecolumns{15}
\tablecaption{TP-AGB candidates identified here and in the literature \label{table:photometry}}

\tablehead{
\multirow{2}{*}{ID} &
\colhead{RA} & 
\colhead{Dec} & 
\colhead{F475W} & 
\colhead{F814W} & 
\colhead{\em J} & 
\colhead{\em K} & 
\colhead{F127M} &
\colhead{F139M} &
\colhead{F153M} &
\multirow{2}{*}{[3.6]} &
\multirow{2}{*}{[4.5]} &
\colhead{Type} &
\colhead{Type} &
\multirow{2}{*}{Type} \\
&
(J2000) &
(J2000) &
{\scriptsize \llap{(0.}48\,$\mu$\rlap{m)}} &
{\scriptsize (0.81)} &
{\scriptsize (1.25)} &
{\scriptsize (2.2)} &
{\scriptsize (1.27)}&
{\scriptsize (1.39)}&
{\scriptsize (1.53)}&
&
&
Lee &
Evans &
}

\startdata
1 & 155.441598 & 18.080108 & 24.73 & 24.09 & \ldots & \ldots & 24.05 & 23.43 & 22.54 & 20.79 & 20.27 & \ldots & \ldots & M \\
2 & 155.434403 & 18.096345 & 24.10 & 20.88 & 19.46 & 18.45 & 19.47 & 19.12 & 19.14 & 18.40 & 18.07 & M & C & C \\
3 & 155.432786 & 18.091945 & 24.18 & 21.03 & 20.07 & 18.77 & 19.69 & 19.29 & 19.40 & 18.19 & 18.14 & C & C & C \\
4 & 155.428871 & 18.077750 & \ldots & \ldots & 22.62 & 19.69 & 23.00 & 21.87 & 21.43 & 18.15 & 18.57 & C & \ldots & C \\
5 & 155.454186 & 18.093011 & \ldots & \ldots & 18.95 & 18.21 & 19.05 & 18.99 & 18.83 & 17.96 & 17.96 & M & \ldots & K/FG \\
6 & 155.426428 & 18.096447 & 24.13 & 21.82 & 20.97 & 20.25 & 20.83 & 20.63 & 20.38 & 19.94 & 19.63 & M & AGB/K & K/FG \\
7 & 155.436872 & 18.092001 & 23.43 & 21.3 & 20.50 & 19.69 & 20.41 & 20.24 & 19.96 & 19.39 & 19.90 & M & AGB & K/FG \\
8 & 155.436079 & 18.082881 & 21.43 & 20.05 & 17.57 & 16.81 & \ldots & \ldots & \ldots & \ldots & \ldots & M & \ldots & K/FG \\
9 & 155.430410 & 18.077854 & \ldots & \ldots & 20.41 & 18.94 & \ldots & \ldots & \ldots & \ldots & \ldots & C & \ldots & Ext \\
10 & 155.423518 & 18.096696 & 23.27 & 21.83 & 21.02 & 19.86 & \ldots & \ldots & \ldots & \ldots & \ldots & M & \ldots & Ext \\
11 & 155.448452 & 18.096050 & \ldots & \ldots & 21.42 & 19.69 & \ldots & \ldots & \ldots & 18.35 & 17.92 & C & \ldots & Ext \\
12 & 155.439832 & 18.091725 & 26.39 & 24.98 & 21.64 & 18.81 & \ldots & \ldots & \ldots & \ldots & \ldots & C & \ldots & ? \\ 
13 & 155.442865 & 18.089298 & 24.67 & 23.93 & 21.59 & 20.10 & \ldots & \ldots & \ldots & \ldots & \ldots & C & \ldots & Ext \\
14 & 155.428950 & 18.102686 & 24.43 & 22.18 & 20.31 & 18.69 & \ldots & \ldots & \ldots & 17.29 & 17.13 & C & \ldots & Ext \\
15 & 155.445400 & 18.097330 & \ldots & \ldots & 21.65 & 20.04 & \ldots & \ldots & \ldots & \ldots & \ldots & C & \ldots & ? \\
16 & 155.447653 & 18.089549 & \ldots & \ldots & 20.81 & 19.33 & \ldots & \ldots & \ldots & \ldots & \ldots & C & \ldots & Ext \\
17 & 155.425790 & 18.081250 & \ldots & \ldots & 20.63 & 19.43 & \ldots & \ldots & \ldots & 18.39 & 18.28 & C & \ldots & Ext \\
18 & 155.438857 & 18.079382 & 24.93 & 23.63 & 21.38 & 19.94 & \ldots & \ldots & \ldots & 18.36 & 17.84 & C & \ldots & Ext \\
19 & 155.427389 & 18.096439 & 24.62 & 22.53 & 21.48 & 20.03 & \ldots & \ldots & \ldots & 19.63 & 19.41 & C & Gal & ? \\
20 & 155.441592 & 18.104558 & \ldots & \ldots & 20.52 & 19.48 & \ldots & \ldots & \ldots & \ldots & \ldots & M & \ldots & ? \\
21 & 155.426829 & 18.100057 & 24.14 & 22.68 & 19.98 & 19.04 & \ldots & \ldots & \ldots & \ldots & \ldots & M & \ldots & Ext \\
22 & 155.436432 & 18.096797 & 23.96 & 22.04 & 21.20 & 20.31 & 21.24 & 21.02 & 20.85 & \ldots & \ldots & M & ? & Ext
\enddata

\tablenotetext{}{\small{\textbf{Note.} The AGB candidates discovered in this work, \citet{Evans2019}, and \citet{Lee2016}. For sources with $HST$ data, coordinates have been corrected using $Gaia$ astrometry ($\Delta \alpha=-0.41^{\prime \prime}, \Delta \delta = -0.53^{\prime \prime}$). For sources without $HST$ data, coordinates are from Gemini \citep{Lee2006}. Photometric data are shown in Vega magnitudes; F475W and F814W are from \citet{McQuinn2015}; {\em J} and {\em K} are from \citet{Lee2016}. IDs are the same as source numbers used within the text. Our chemical classifications as well as those determined by \citet{Lee2016} and \citet{Evans2019} are shown in the last three columns. Sources with classifications but no photometry were cut in either the crowding, sharpness, or TRGB cuts. Sources labeled as K/FG were classified as likely K-type giants or foreground stars; sources labeled Ext are classified as extended sources and likely background galaxies.} \vspace{-0.5cm}}
\end{deluxetable*}

\section{Observations and Methods}
\label{sec:observations} 

Observations included imaging using both the \textit{Spitzer Space Telescope} and \textit{HST} to isolate AGB stars by their chemical type in the near-IR, and determine whether they are producing significant dust in the IR. Table \ref{table:observations} provides the details of the observations. The {\it HST} observations occurred just ten days after the {\it Spitzer} observations, minimizing issues with stellar variability, since TP-AGB stars can vary by $>$1~mag at near-IR wavelengths over $\sim$\,300--1000~days. Table \ref{table:photometry} shows the photometry from our observations as well as archival observations. The \textit{HST} observations were sensitive enough to detect stars 3 magnitudes below the TRGB, and the sensitivity of our \textit{Spitzer} observations begins to fall off after [4.5] $\sim$\,18 mag (Figure \ref{fig: luminosity function}). \vspace{0.5cm}

\subsection{Spitzer Space Telescope Imaging}
Leo P was observed using \textit{Spitzer} in the IRAC bands at 3.6 and 4.5\,$\mu$m. We have performed point-spread function (PSF) photometry on the dithered exposures using the DUSTiNGS pipeline \citep{Boyer2015a}, which uses {\asciifamily DAOphot II} and {\asciifamily ALLSTAR} \citep{Stetson1987}. The observations consisted of four exposures using the 16-point spiral medium dither pattern. For one source (Source 1; described further in \S4), the 3.6~$\mu$m counterpart was not originally detected by the DUSTiNGS photometry pipeline.  To overcome this, we performed aperture photometry with a 2$^{\prime\prime}$ aperture on both the 3.6 and 4.5~$\mu$m images to determine the [3.6]-[4.5] color ($\approx$\,0.5~mag). We then applied this color to the PSF-derived 4.5~$\mu$m magnitude to obtain a 3.6~$\mu$m magnitude.

\subsection{Hubble Space Telescope Imaging}

Leo\,P was imaged in the near-infrared with {\it HST's} Wide Field Camera 3 infrared channel (WFC3/IR), using the F127M, F139M, and F153M filters with pivot wavelengths of 1.27, 1.38, and 1.53\,$\mu$m, respectively\footnote{The DOI for this dataset as well as the wide-band $ACS$:$HST$ optical data from \citet{McQuinn2015} is \href{http://dx.doi.org/10.17909/t9-67a4-t896}{10.17909/t9-67a4-t896}.} (program GO-14845). These filters allow for the disentanglement of the carbon- and oxygen-rich AGB stars \citep[e.g.][]{Boyer2017}. A four-position box dither pattern was employed to mitigate imaging artifacts and increase the spatial resolution, with a total exposure time of $\sim$800~s for F127M and F153M and $\sim$850~s for F139M, which has a slightly lower throughput. The detector was read out using the WFC3/IR {\asciifamily STEP} sequence, which samples the detector with the shortest possible sequence for the first few reads, then samples logarithmically thereafter to provide a high dynamic range for fields containing both bright and faint sources. Figure \ref{hst_image} shows a mosaic image using the F127M filter data.

We performed PSF photometry using {\asciifamily DOLPHOT's} WFC3-specific module \citep{Dolphin2000}. A drizzled mosaic of the F127M images was used as the reference image (created using \textit{HST} {\asciifamily Drizzlepac} v2.0) to allow forced photometry of faint objects in the individual images. To limit contamination from background galaxies and foreground sources, we have implemented cuts for sharpness and crowding in all three bands: ($\Sigma$\,Sharp$_{\lambda}$)$^2 < 0.5$ and ($\Sigma$\,Crowd$_{\lambda}$)$^2 < 5$ plus an additional constraint in the F127M band: Sharp\textsubscript{F127M} $>$ $-$0.025 and Crowd\textsubscript{F127M} $<$ 1.5. By restricting the sharpness parameter, we minimize contamination from extended objects and cosmic rays. The crowding restriction eliminates objects that are significantly affected by nearby objects. We visually inspected all objects of interest to confirm that they are truly unblended point sources.

The measured reddening towards Leo P is $E(B-V)$\,=\,$0.0227$\,mag \citep{Schlafly2011}, and assuming an $R \textsubscript{V}$\,=\,$3.1$, this corresponds to an extinction $A\textsubscript{V}$\,=\,$0.0693$ mag. The photometry in each of the \hst filters has been corrected for this interstellar extinction using the corrections from \citet{Cardelli1989}, an approximate shift of $-0.016$ mag. The \textit{Spitzer} data are uncorrected because the amount of extinction is smaller than the photometric uncertainties in the IRAC filters.

\subsection{Previous observations}

Previous studies have used the Gemini North Telescope and the European Southern Observatory's Very Large Telescope (VLT) to search for evolved stars in Leo P. These programs have detected eleven oxygen-rich AGB candidates and fifteen carbon-rich AGB candidates \citep{Lee2016,Evans2019}.

\paragraph{Gemini North-NIRI}

\citet{Lee2016} used the Near InfraRed Imager and Spectrometer \citep[NIRI;][]{Hodapp2003} on the Gemini North Telescope to target AGB stars using the $J$ and $K$ bands and determine their chemical type. The $J-K$ color probes circumstellar extinction. These bands are not, however, as effective for determining the chemistry of more evolved dusty TP-AGB stars, as molecular features within the $J$ waveband make AGB stars redder in their $J-K$ color. \citet{Lee2016} identified nine oxygen-rich and thirteen carbon-rich AGB candidates.

\paragraph{VLT-MUSE}
\citet{Evans2019} obtained optical spectra with the VLT's Multi Unit Spectroscopic Explorer \citep[MUSE;][]{Bacon2010}. These observations cover the 483--930\,nm wavelength range, which includes TiO and C$_2$ molecular bands and the Ca\,\textsc{ii} triplet. C$_2$ Swan bandheads at 5165 and 5636\,{\AA} were detected in two AGB candidates, confirming their carbon-rich chemical type. Two additional sources show the Ca\,\textsc{ii} triplet in absorption, where it was used to measure a recessional heliocentric velocity in one of these sources (Source 7; 262\,$\pm$\,8 km\,s$^{-1}$) that was consistent with membership of Leo P. The second source (Source 22) was only detected in one Ca\,\textsc{ii} band, and may be either a foreground star, K-type giant, or an AGB star in Leo P.

\subsection{Determining AGB chemistry and isolating dusty AGB candidates}
In order to isolate the evolved AGB population of Leo P, we have limited our sample using cuts in magnitude and F127M$-$F153M color. The dust surrounding TP-AGB stars obscures them in optical bands and increases their F127M$-$F153M color. We have thus only considered sources with F127M$-$F153M color typically redder than the Red Giant Branch (RGB) following the reddening vector, or sources above the Tip of the Red Giant Branch \citep[TRGB; M\textsubscript{F127M}$ < 21$\,mag;][]{Boyer2017} as possible AGB candidates. We also considered one source to the left of the RGB where oxygen-rich stars have been shown to show deep water absorption \citep[Figure 5 in ][]{Boyer2019}.

\begin{figure*}
 \centering 
 \includegraphics[height=8.87cm]{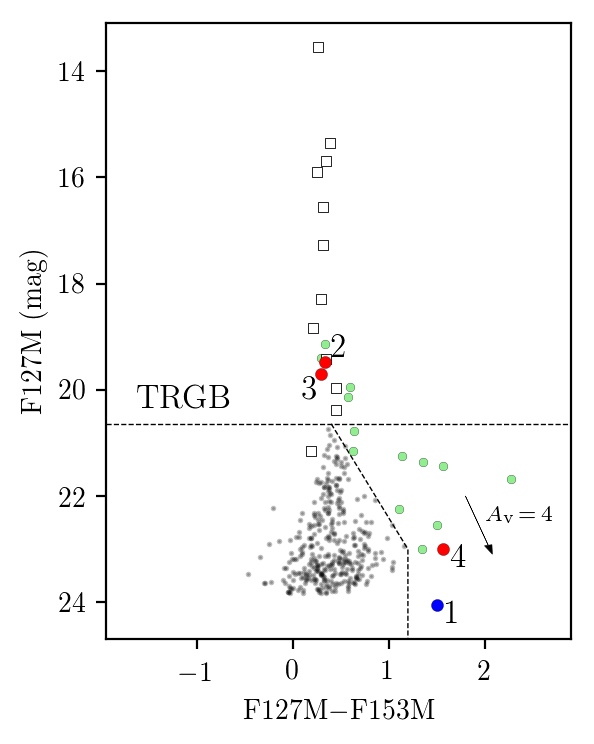}
 \includegraphics[height=8.87cm]{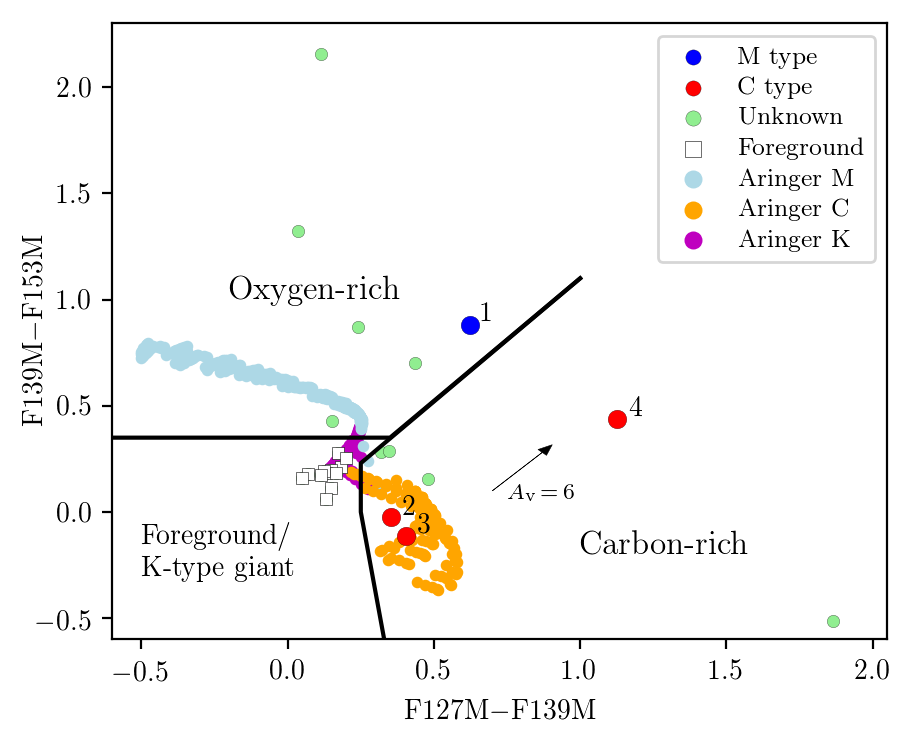}
 \caption{\hst color-magnitude and color-color diagrams showing (\textit{Left}) sources above the Tip of the Red Giant Branch (TRGB) or dusty enough to be considered AGB candidates and (\textit{Right}) a utilization of the method successfully used by \citet{Boyer2013,Boyer2017,Boyer2019} to disentangle carbon- (C-type; red) and oxygen-rich (M-type; blue) evolved stars, and sources classified as either foreground stars or K-type giants (white squares); colors apply to both diagrams. Also shown are stellar models from \citet{Aringer2009} and \citet{Aringer2016} as well as extinction vectors showing reasonable values for $A_{\rm V}$, which assume an $R_{\rm V}$\,=\,3.1 and the extinction law from \citet{Cardelli1989}. While the extinction law applies to interstellar dust, this level of obscuration will have a similar effect on AGB dust. Sources that meet our TP-AGB selection criteria, but are visibly extended or not point-source-like in the \textit{HST} images (shown in Appendix A), are labeled as ``Unknown'' (green). \\} 
 \label{fig: trgb_knee} 
\end{figure*} 

\citet{Boyer2013,Boyer2017} showed that the \hst WFC3/IR F127M, F139M, and F153M filters separate AGB stars by their chemical type \citep[Figure 1 from][]{Boyer2017}. These filters probe wavebands that are affected by different molecular features; H$_2$O for oxygen-rich AGB stars and CN+C$_2$ for carbon stars. Alongside the stellar evolutionary models, in the F127M$-$F139M and F139M$-$F153M color-color space (Figure \ref{fig: trgb_knee}), our sources clearly separate into two populations; oxygen- and carbon-rich.

\begin{figure*}
 \centering
 \includegraphics[width=0.246\linewidth]{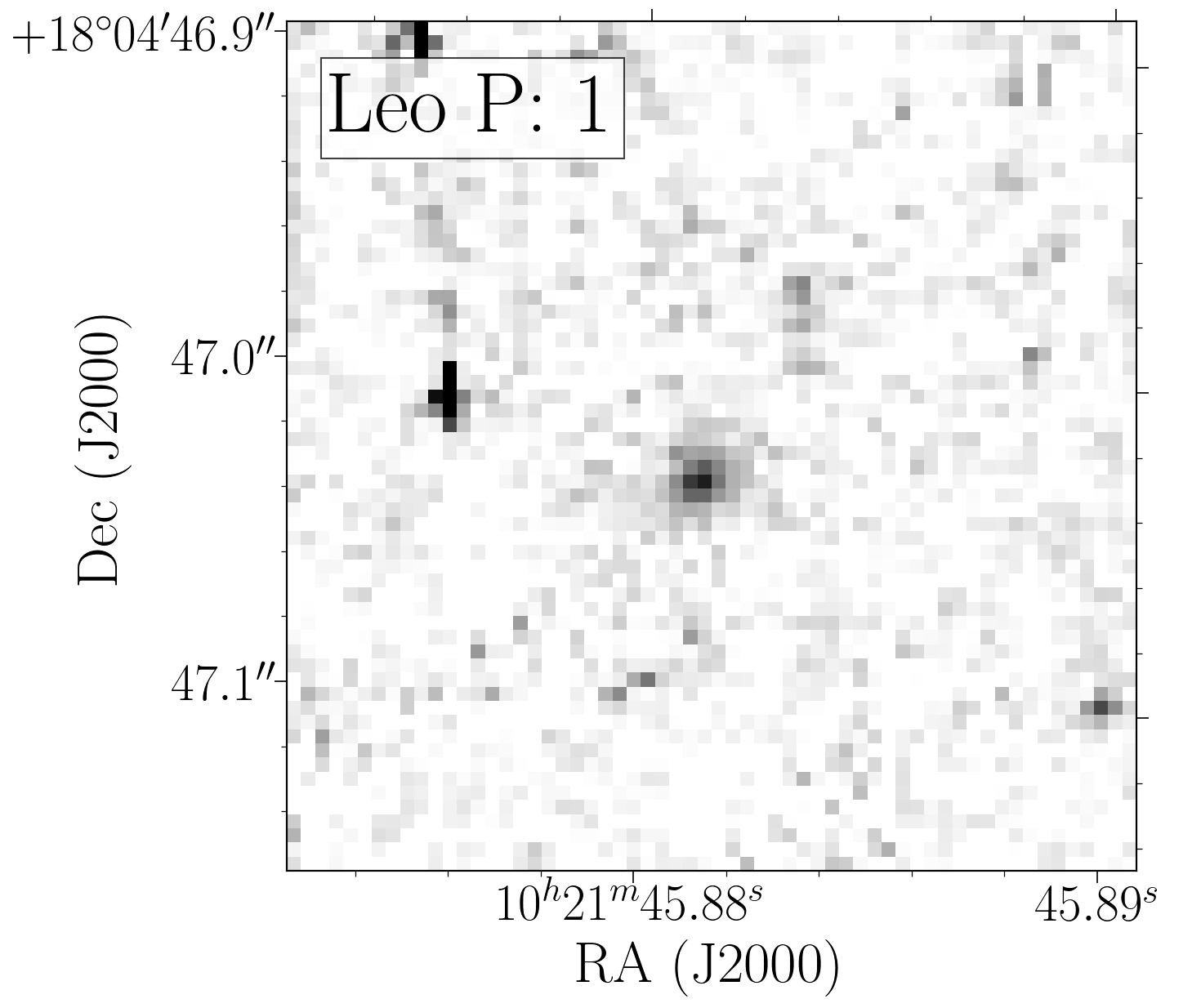}
 \includegraphics[width=0.246\linewidth]{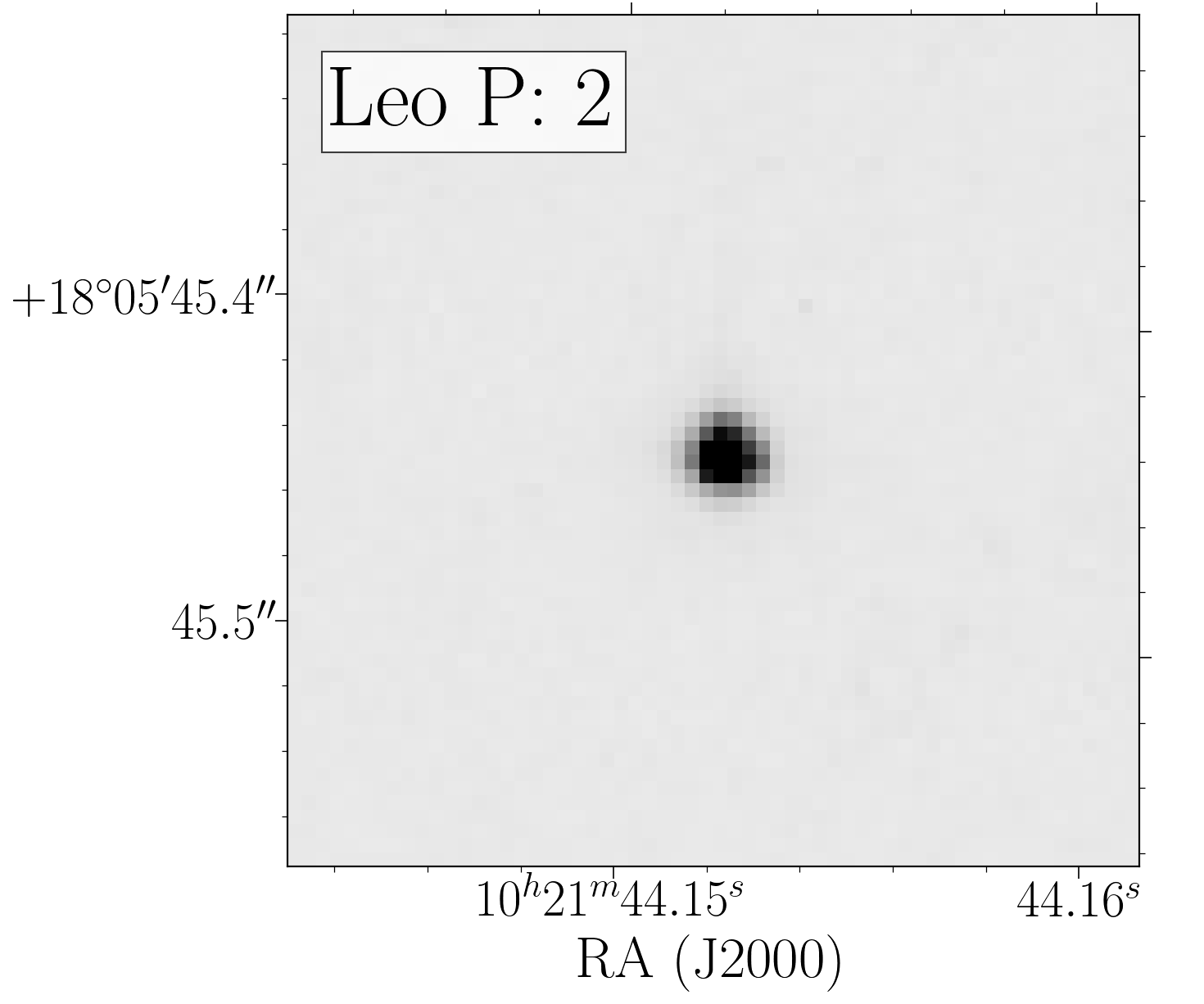}
 \includegraphics[width=0.246\linewidth]{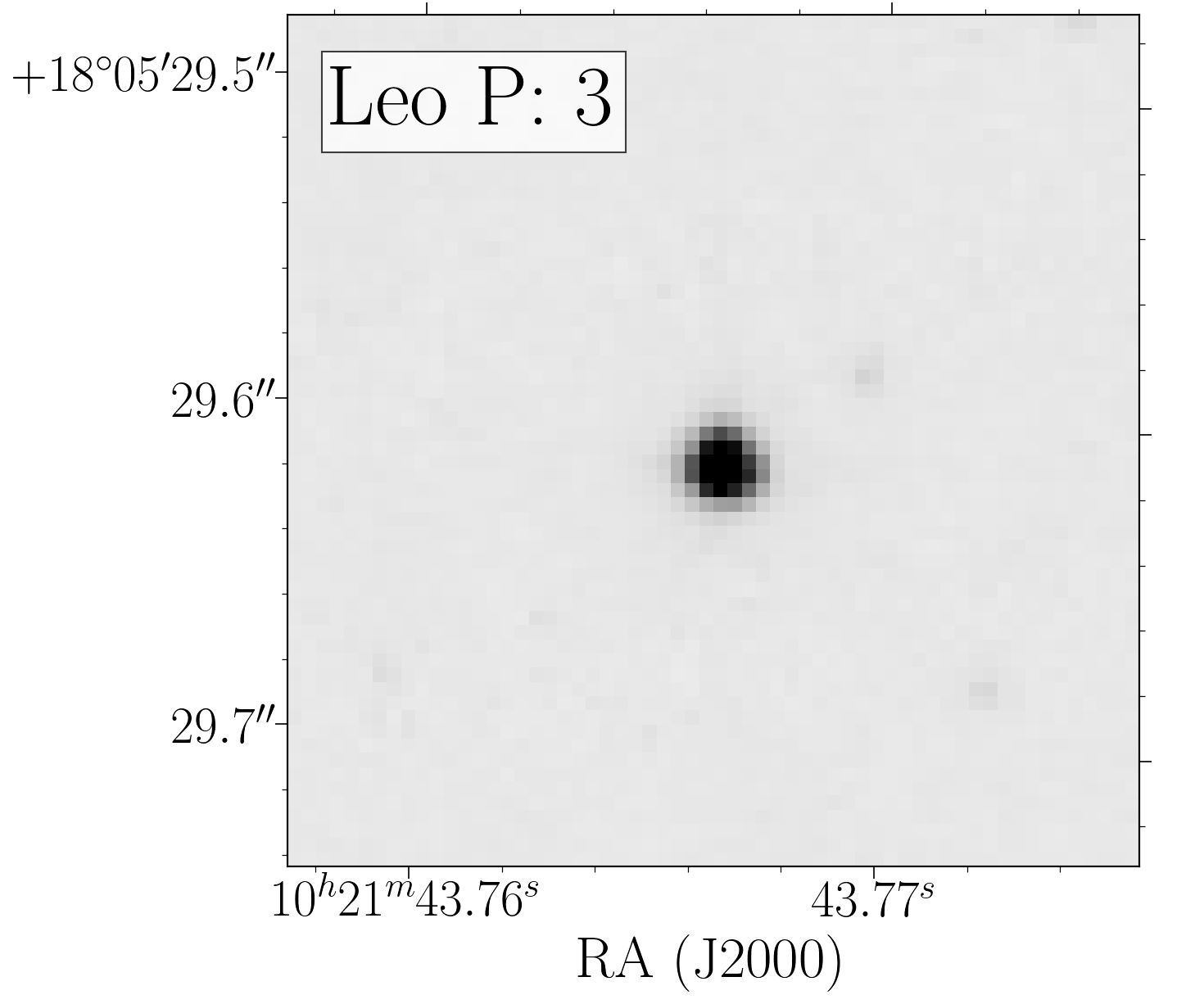}
 \includegraphics[width=0.246\linewidth]{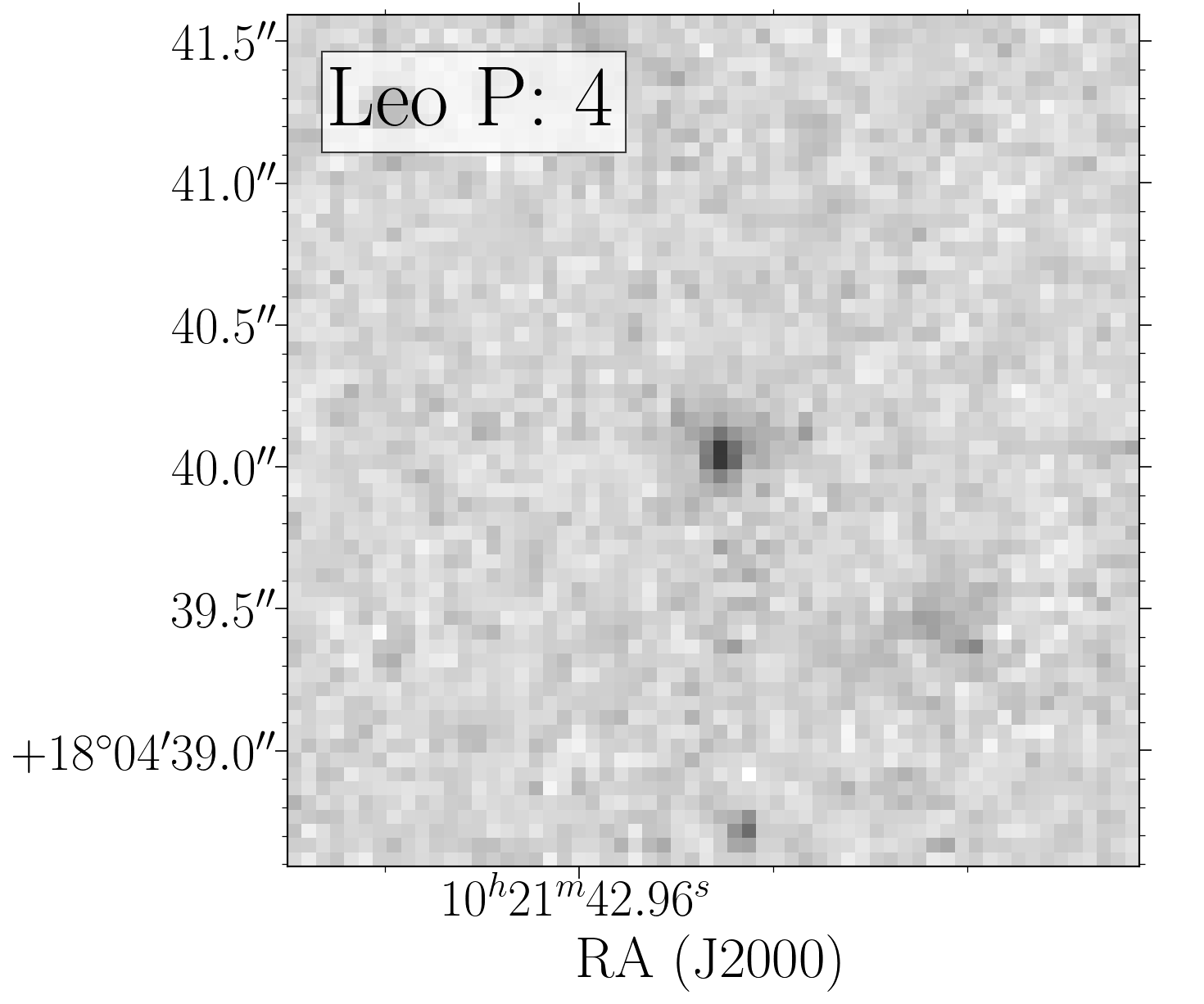}
 \caption{\textit{HST}:\textit{ACS} F814W mosaic cutouts of our four AGB candidates. All four sources are point-like and unlikely to be background galaxies. Source 1 (\textit{Left}) and  Source 4 (\textit{Right}) are expected to be dusty. \\} 
 \label{fig:gold_stamps}
\end{figure*}

\section{Results and Discussion}
We have successfully identified four AGB candidates and classified them by chemical type. We have also shown that the method of AGB classification of carbon stars using $HST$ medium-band filters is robust even at extremely low metallicities; this remains to be confirmed in our dusty AGB candidates. Table \ref{table:photometry} shows photometry from \textit{HST}, \textit{Spitzer}, and NIRI, as well as our classification and the classifications by \citet{Lee2016} and \citet{Evans2019}. Sources with classifications but no \textit{HST} photometry were cut in the crowding, sharpness, or TRGB cuts, and are not likely to be AGB stars (see \S2). Figure \ref{fig:gold_stamps} shows \textit{HST/ACS} F814W mosaic snapshots of our AGB candidates. Our sample of  AGB candidates is similar in number to those found in other galaxies of similar size to Leo P ($M_V=-9.3$\,mag), for example Phoenix ($M_V=-9.9$\,mag) or Sag DIG ($M_V=-11.5$\,mag) with one and three sources, respectively \citep{Menzies2008,McConnachie2012,Whitelock2018}. Of the four sources detected, two appear dusty. This is consistent with the estimate number of AGB stars predicted using {\asciifamily MATCH} ($N_{\rm AGB} \sim 3$).\\

\subsection{Contaminant sources}
\label{sect:contamination}
Searches for evolved stars in the near-IR and optical are commonly affected by contaminating sources, typically foreground stars or K-type giants. With the angular resolution of \hst we are able to (and using sharpness cutoffs) identify and remove any extended sources such as background galaxies visually. For the point-like sources, we can identify stars that may be either foreground stars or warmer giants and supergiants by their position in color-color space (shown in Table \ref{table:garbage_photometry} and Figures \ref{fig: trgb_knee} \& \ref{fig: stamps}); results from the {\asciifamily Trilegal} simulations have confirmed foreground stars with these near-IR colors \citep{Boyer2017}. Point sources may also be young stellar objects (YSOs), PNe, or post-AGB stars. These phases are typically much shorter than the AGB, and thus we expect contamination from these sources to be rare. Of the twenty-five sources that passed the photometry and TRGB cutoffs, we have classified four as AGB candidates, nine as extended sources, and twelve as either foreground stars or K-giants. This was done by in part by inspecting the shape of the source within the image. Without high-resolution imaging, many of the extended background galaxies would have been classified as evolved stars. Confirmation of these classifications as well as our AGB candidates will require spectroscopic follow-up.

Red supergiants (RSGs) could possibly also contaminate our sample. These sources, while rarer than AGB stars, may be mistakenly classified as luminous oxygen-rich AGB candidates. Within the SMC at one-fifth solar metallicity, RSGs have been shown to have warmer median temperatures \citep{Levesque2006,Massey2007}. It is also expected that a decrease in metal content will result in a decrease in opacity and thus an increase in surface brightness. Given these assumptions, any RSGs should lie closer to the foreground sources in our color-color diagram (Figure \ref{fig: trgb_knee}). Given the low brightness of our oxygen-rich candidate, we do not expect it to be a RSG.

\begin{figure}  
 \centering
 \includegraphics[width=\linewidth]{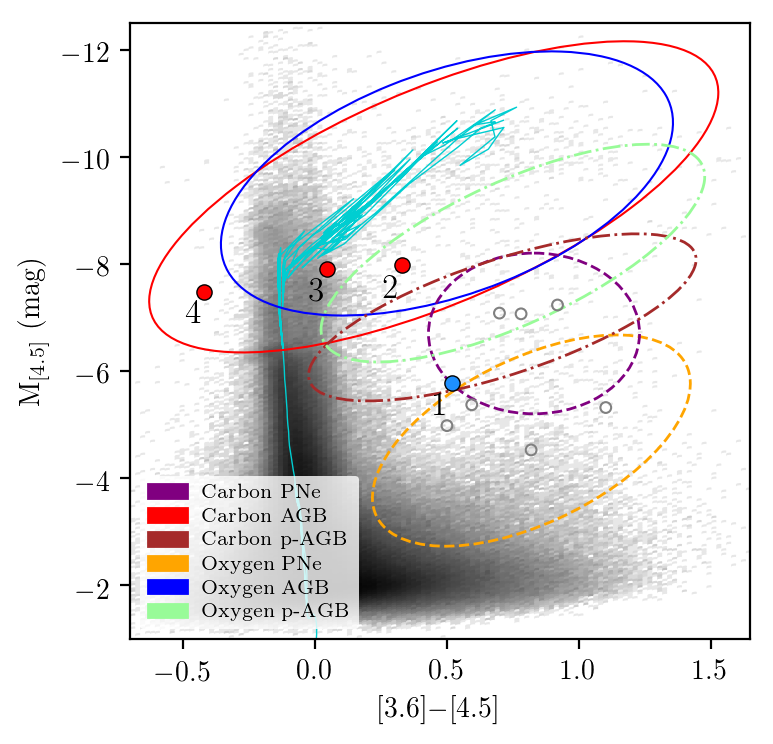}
 \caption{An IRAC CMD showing our oxygen (blue circle) and carbon (red circles) AGB candidates in Leo P, along with the locations of other spectroscopically confirmed classes of LMC stars from SAGE-Spec \citep{Jones2017}. Shown are AGB stars (solid lines), PNe (dashed lines), post-AGB (dot-dashed lines), background galaxies (gray circles), and remaining SAGE-SPEC sources (background points). A COLIBRI isochrone using the \citet{Nanni2013,Nanni2014,Nanni2016} dust growth models with a log(age in years) = 8.8, LMC metallicity, and a normalized number of seed particles log${\epsilon}_{\rm S}$ = $-$13, is also shown in cyan. \\}
 \label{fig: cmd} 
\end{figure}

\subsection{Oxygen-rich AGB candidate}
\label{sect:o-rich_candidate} 
\begin{figure}
 \centering \hspace{0.1cm}
 \includegraphics[width=0.99\linewidth]{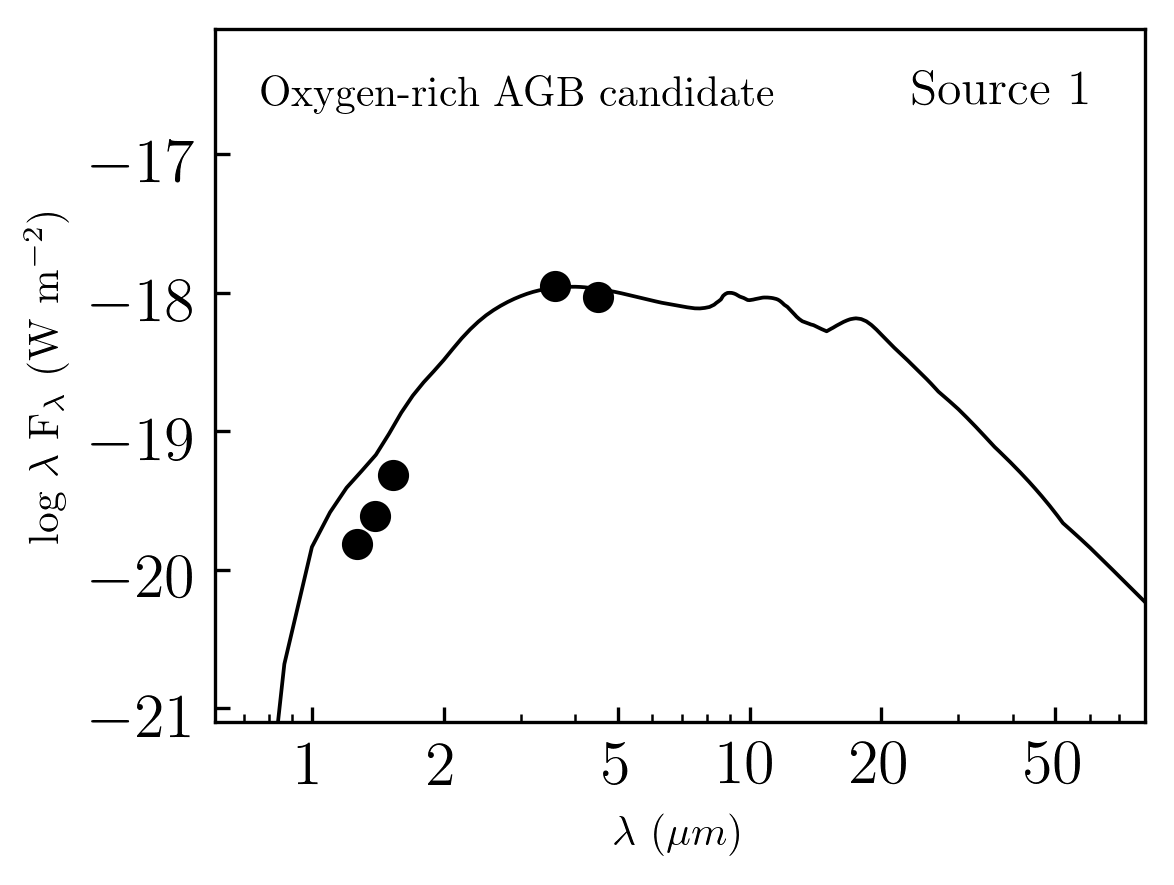}
 \caption{The Spectral Energy Distribution of the oxygen-rich AGB candidate, shown with {\it Spitzer} and {\it HST} photometry, as well as a {\asciifamily DUSTY} radiative transfer model \citep{Elitzur2001} that uses a black body and grains from \citet{Ossenkopf1992}, fitted using the {\asciifamily Dusty Evolved Star Kit} ({\asciifamily DESK}).\\}
 \label{fig: o_sed_fit} 
\end{figure}

We have detected one oxygen-rich AGB candidate in Leo P (Source 1). Figures \ref{fig: trgb_knee} and \ref{fig: cmd} show the source's position in near-IR CMDs. While we expect that the star is obscured and fainter in the \textit{HST} wavebands as a result of dust, the \textit{Spitzer} photometry ($[4.5] = 20.27$\,mag) is also fainter than what is expected for a TP-AGB star. The source lies around two magnitudes below the oxygen-rich AGB stars found in the SAGE-LMC sample at 4.5\,$\mu$m. We used the {\asciifamily Dusty Evolved Star Kit}\footnote{\url{https://github.com/s-goldman/Dusty-Evolved-Star-Kit}} ({\asciifamily DESK}), a \textsc{python} package for fitting the SEDs of AGB stars with a range of model grids using a least-squares fit. The source was fitted with a grid of {\asciifamily DUSTY} models that use grains from \citet{Ossenkopf1992}. These models, however, do not fit all of the \textit{HST} filters well (Figure \ref{fig: o_sed_fit}). The source is unlikely to be a YSO as it is far from any nearby star-forming regions. This source may not be a massive O-rich TP-AGB star, but rather a post-AGB star or PN. If this is the case, the source may still be showing that silicate dust can form at these metallicities. Another possibility is that the source is a background galaxy, however, the full eidth at half maximum (FWHM) of the source is within the range of other confirmed nearby stars. Spectroscopic follow-up is needed to confirm the nature of this source.

\subsection{Carbon-rich AGB candidates}
\begin{figure}
 \centering \hspace{0.1cm}
 \includegraphics[width=0.99\linewidth]{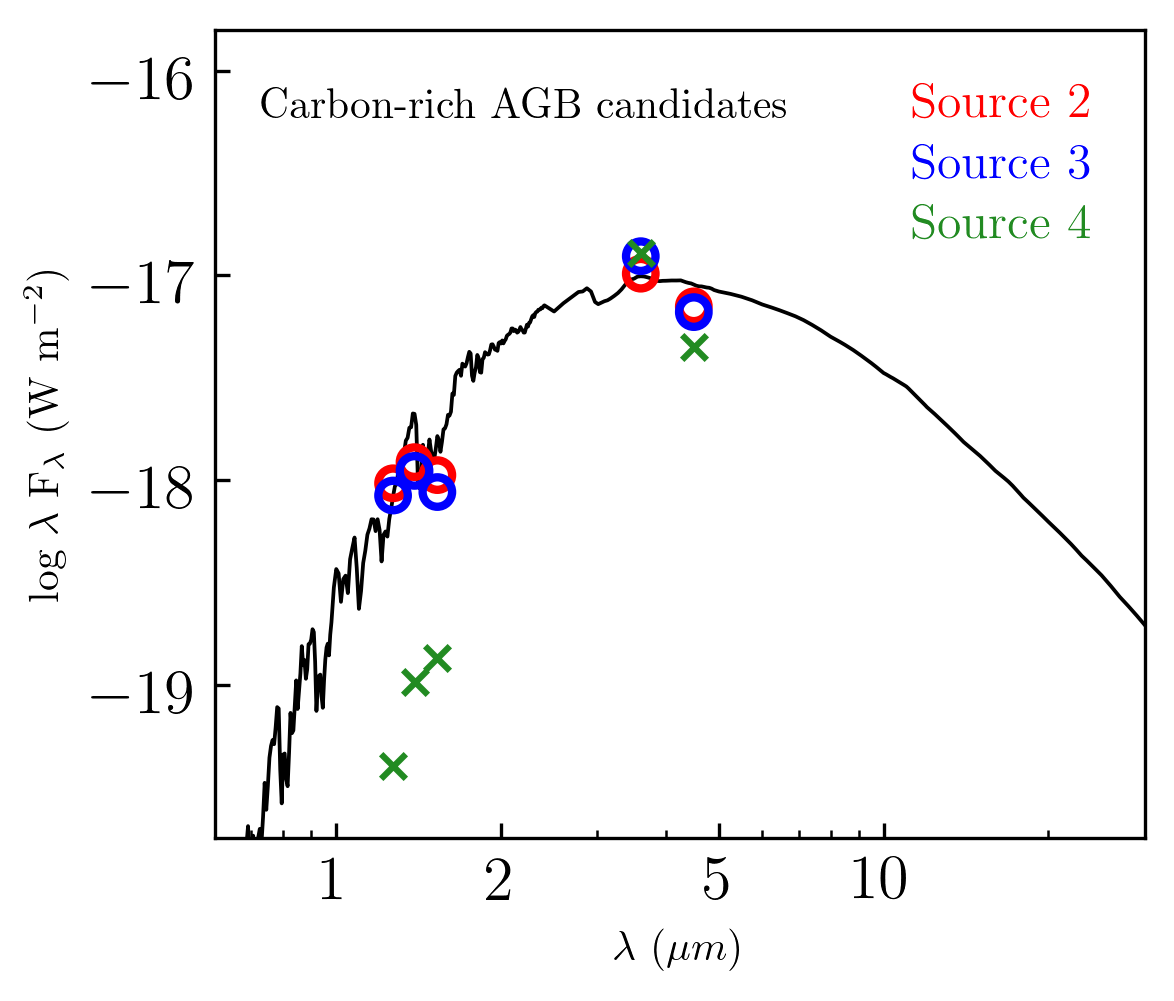}
 \caption{The SEDs of the carbon-rich AGB candidates, shown with {\it Spitzer} and {\it HST} photometry, as well as models from the dust growth code from \citet{Nanni2018}. Shown is the best-fitting model of Source 3 using a LMC carbon model (solid black) which uses dust grains from \citet{Hanner1988}.\\}
 \label{fig: c_sed_fit}
\end{figure}

We have detected three carbon-rich AGB candidates in Leo P, including one that may be dusty. The two sources that appear dustless based on their \textit{Spitzer} and \textit{HST} colors (Sources 2 and 3) lie above the TRGB and were also previously categorized as carbon-rich AGB candidates by their strong C$_2$ Swan bandheads at 5165 and 5636\,{\AA} \citep{Evans2019}. Figure \ref{fig: c_sed_fit} shows the spectral energy distribution (SED) of Sources 2, 3, and 4, along with radiative transfer models from the dust-growth code of \citet{Nanni2018}, fit with the {\asciifamily DESK}. Shown is a best-fitting model of Source 3. Dustier models fit poorly, because the dust shifts too much emission into the IR. The photospheric CO absorption and poor fit with dusty models suggest that Sources 2 \& 3 are a carbon-rich AGB star with little to no dust.

The dusty candidate (Source 4) has a uniquely-shaped SED that begins to fall in brightness longwards of 4\,$\mu$m (Figure \ref{fig: c_sed_fit}). Carbon stars tend to have bluer [3.6]$-$[4.5] colors as a result of CO absorption at 4--6\,$\mu$m. This feature gets veiled with increasing dust \citep{Frogel1972,Boyer2011,Blum2014}. Source 4 has a [3.6]$-$[4.5] color of $-$0.42 mag, indicative of CO absorption. The medium-band \textit{HST} colors also suggest a source that is heavily obscured by dust (F127M$-$F139M = 1.12 mag). It may be the case that the source is not dusty enough to be veiled by the continuum dust emission, yet still shows signs of heavy dust obscuration in the optical. Another alternative is that this source may have a unique geometry that reveals both the dust emission and photosphere of the source. 

It is also possible that Source 4 is not an AGB star but an entirely different phase of evolution. It is located outside of the half-light radius of the galaxy (Figure \ref{hst_image}). While AGB stars have previously been discovered in the outskirts of dwarf galaxies \citep{McQuinn2017}, fewer of them are expected at these distances. AGB stars at these distances may possibly indicate a ``bursty'' SFH that causes star formation to occur more in the outskirts of the galaxy at later times. This would then create younger populations to mix with older populations \citep[inside-out galaxy formation;][]{El-Badry2016}. 

\subsection{Comparison with previous detections}

We have independently recovered the two carbon-rich AGB candidates previously identified spectroscopically (Sources 2 and 3). Our two new dusty AGB candidates (Sources 1 and 4) were found outside of the footprint of the previous MUSE observations and may have been too faint to be detected in the broad-band near-IR imaging with NIRI. 

We have also detected the two oxygen-rich AGB candidates found by \citet[][Sources 7 and 22]{Evans2019}, which fall in the ``Foreground / K-type giant'' area of our color-color diagram (Figure \ref{fig: trgb_knee}). While only one of these sources has a confirmed membership to Leo P (Source 22), neither source was found to have significant TiO bands in their optical spectra, and we do not see evidence of H$_2$O in our more recent observations. In metal-poor environments it is expected that the lower natal abundance of oxygen results in warmer atmospheres and a higher fraction of K-type giants that do not produce dust. 

The additional candidates only found by \citet{Lee2016} may be highlighting the limitations of using broad-band and low-resolution optical photometry to identify and classify AGB stars. These bands are affected by molecular absorption as well as both carbon- and oxygen-rich dust, making these bands less reliable for categorizing and classifying dusty AGB stars. We have determined that ten of their 21 candidates are extended, and four are either foreground stars or K-type giants.

\subsection{Leo P as an analog of a high-redshift galaxy}

Leo P's isolation makes it a useful analog for high-redshift galaxies. Being unaffected by galaxy interactions or stripping events (ram-pressure or tidal) ensures that we have a better understanding of its history. Due to their longer evolutionary time scales, low-mass AGB stars may be limited in their number in high-redshift galaxies, as opposed to Leo P. Galaxies out to a redshift of $\sim$\,8, however, will have had enough time for both oxygen- and carbon-rich AGB stars to form and produce dust \citep[][and references therein]{Sloan2009}. As Leo P is less massive, it is also expected to have a smaller stellar population, making it more difficult to catch shorter phases of evolution like the AGB. 

In terms of their chemical abundances, Leo P has only retained 5\% of its SNe ejecta \citep{McQuinn2015a}, limiting the abundance of r-process elements, which can be important sites of dust nucleation. While carbon stars are expected in both environments, the difference in the amount of carbon present may differ. Additional carbon in either environment may lock up more oxygen in CO, limiting the ability of the oxygen-rich AGB stars to produce dust. The strength of this effect at these metallicities, however, is not well known. Understanding the impact of metallicity on dust production on the AGB will require observing additional metal-poor samples in galaxies like I Zw 18 or Leoncino \citep[AGC 198691;][]{Hirschauer2016}.

\section{Conclusion}
\label{sec:conclusion}

We have conducted a census of the population of evolved stars in the metal-poor dwarf galaxy Leo P. This galaxy is the most metal-poor gas-rich galaxy that we can resolve with current instruments and is an excellent environment for studying stellar populations similar to those in high-redshift galaxies. Our census has yielded four AGB candidates. They were selected on the basis of their near-IR medium-band photometry. We have found most of the AGB candidates previously detected by low-resolution imaging to be extended objects and likely background galaxies. 

The one oxygen-rich candidate and one of the carbon-rich candidates appear to be dusty in both the \textit{Spitzer} and \textit{HST} data, while the other two carbon-rich candidates appear dust-free. Our oxygen-rich AGB candidate shows high levels of obscuration in the \textit{HST} bands, and a reddened color in the \textit{Spitzer} bands, indicative of dust. The low luminosity in these bands may be the result of metallicity effects or changes in dust properties, but is more likely to suggest that the source is in another phase of evolution (e.g., post-AGB, PN). Given the small mass of Leo P and the short length of these phases, this source is exceptionally rare. If the source is a dust-producing AGB star, however, it would likely dominate the dust production within Leo P. The dusty carbon-rich candidate appears to be highly obscured in the \textit{HST} data ([1.27]$-$[1.53]\,=\,1.55\,mag), but does not show reddened \textit{Spitzer} colors. The source has a blue [3.6]$-$[4.5] color of $-$0.42\,mag, likely from strong CO absorption at 4.5\,$\mu$m, which is commonly found in carbon stars. The heavy extinction in the \textit{HST} bands may also arise from a unique geometry.

These candidates may allow us to better understand dust production in metal-poor environments and high-redshift galaxies, but this requires follow-up in the IR. Spectroscopic follow-up with the $JWST$ will allow us to both confirm the nature of these sources and study how dust forms at these metallicities ($Z$\,$\sim$\,2$\%$\,Z$_{\odot}$). The dust budget of a galaxy of this size is also likely dominated by only a few sources, highlighting the need to understand how efficiently they produce dust, as well as the composition of the dust that is produced.

\section*{Acknowledgments}

Support for program HST-GO-14845 was provided by NASA through a grant from the Space Telescope Science Institute. STScI is operated by the Association of Universities for Research in Astronomy, Inc. under NASA contract NAS 5-26555. This work is based in part on observations made with the Spitzer Space Telescope, which is operated by the Jet Propulsion Laboratory, California Institute of Technology under a contract with NASA. IM and AAZ acknowledge support from the UK Science and Technology Facility Council under grants ST/L000768/1 and ST/P000649/1. This research has also made use of the Vizier catalogue access tool, CDS, Strasbourg, France. This research has made use of the Dusty Evolved Star Kit (DESK; \href{https://github.com/s-goldman/Dusty-Evolved-Star-Kit}{ https://github.com/s-goldman/Dusty-Evolved-Star-Kit}).

\facility{\it Spitzer Space Telescope, Hubble Space Telescope}

\software{DESK, MATCH \citep{Dolphin2002}, DUSTiNGS pipeline \citep{Boyer2015a}, DOLPHOT \citep{Dolphin2000}}

\bibliographystyle{aasjournal.bst}
{\small
\bibliography{references_2019}}

\appendix 
\section{Galaxies} After classifying our AGB candidates using color-color cuts, the \hst image revealed that several sources were not clear point sources. These are sources above the TRGB, but that were determined to be either K-type giants or foreground stars (K/FG), extended sources or background galaxies (Ext), or sources that did not meet our sharpness or crowding cutoffs.

\begin{figure*}[h]

 \includegraphics[width=0.32\linewidth]{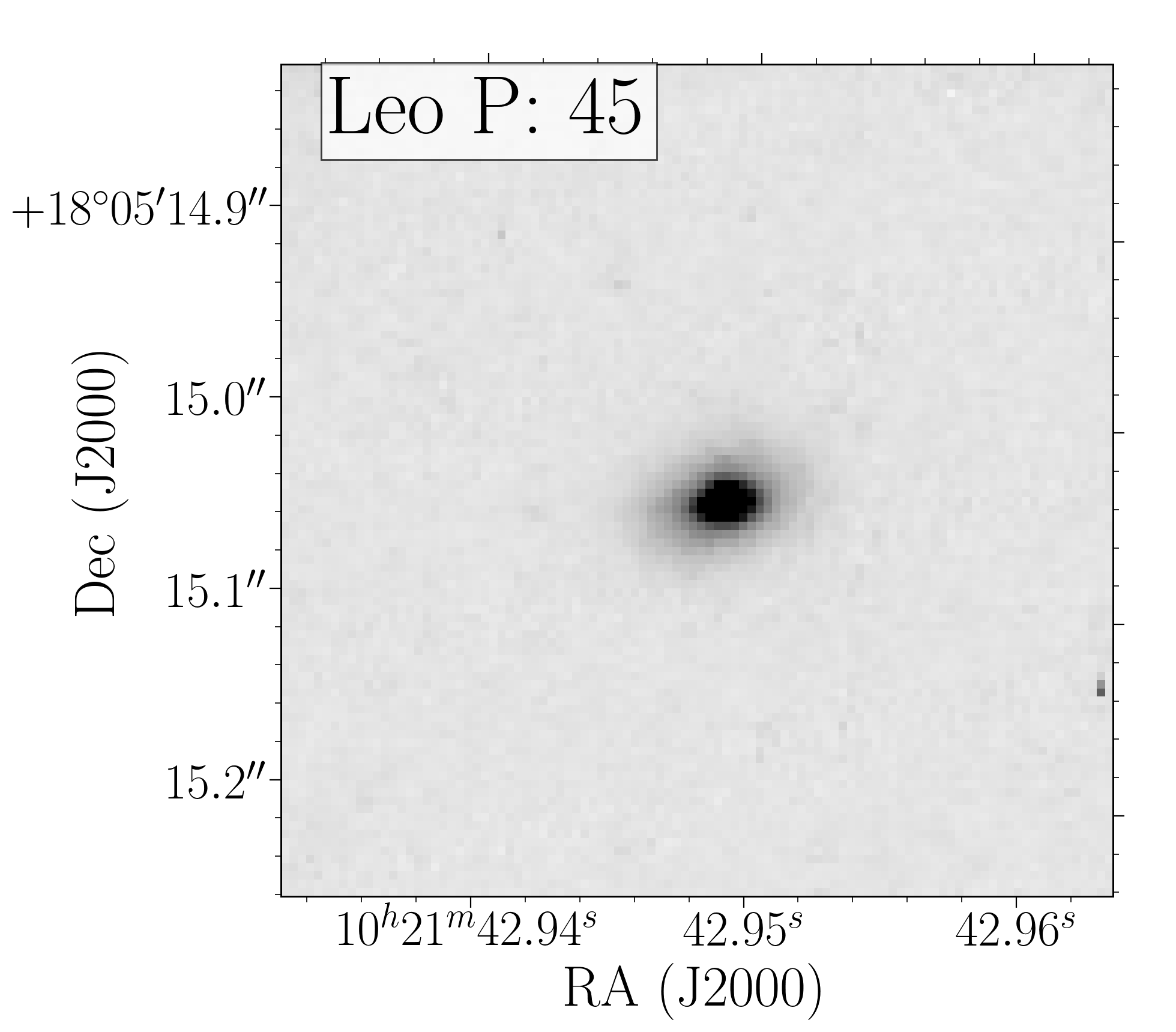}
 \includegraphics[width=0.32\linewidth]{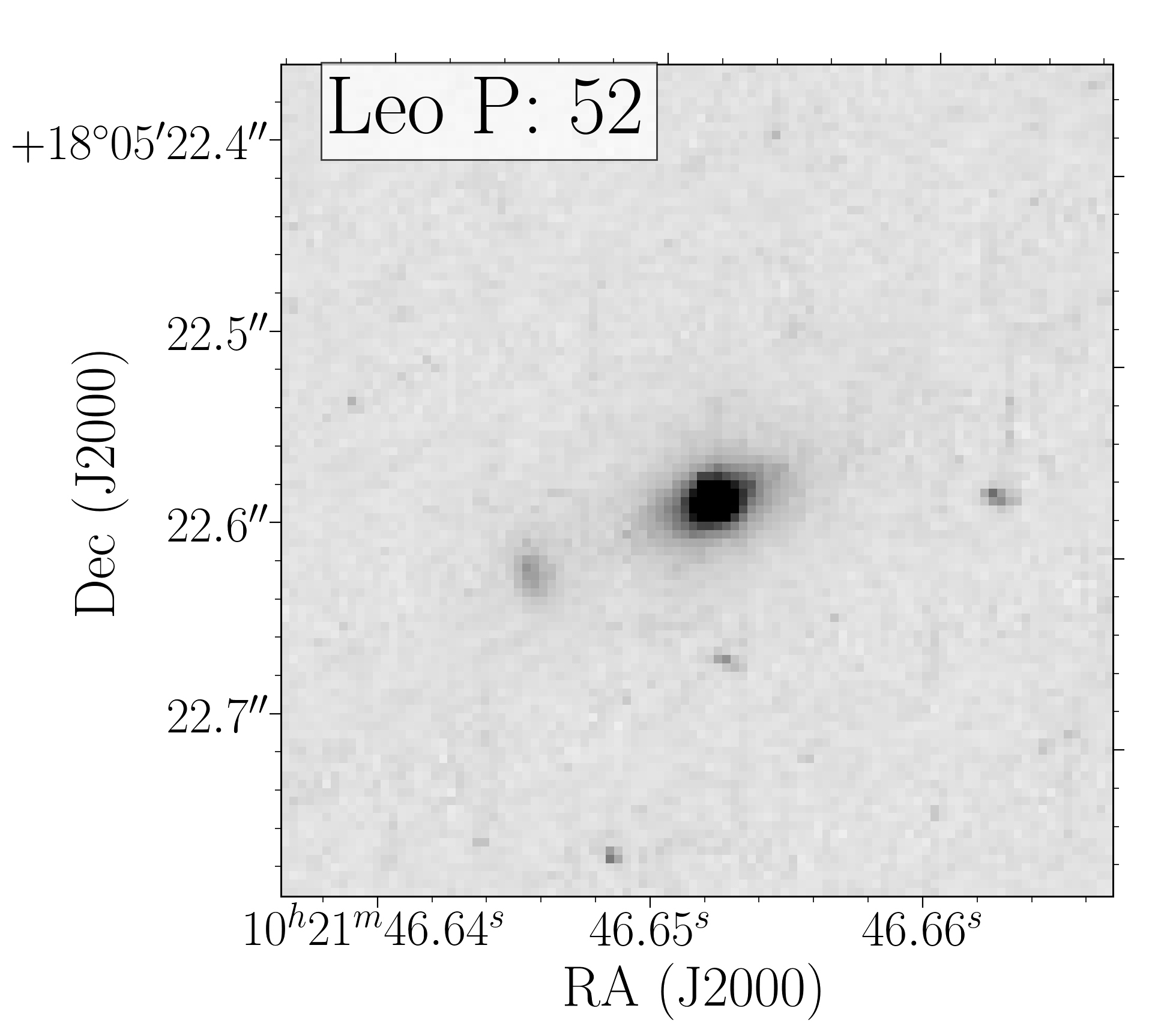}
 \includegraphics[width=0.32\linewidth]{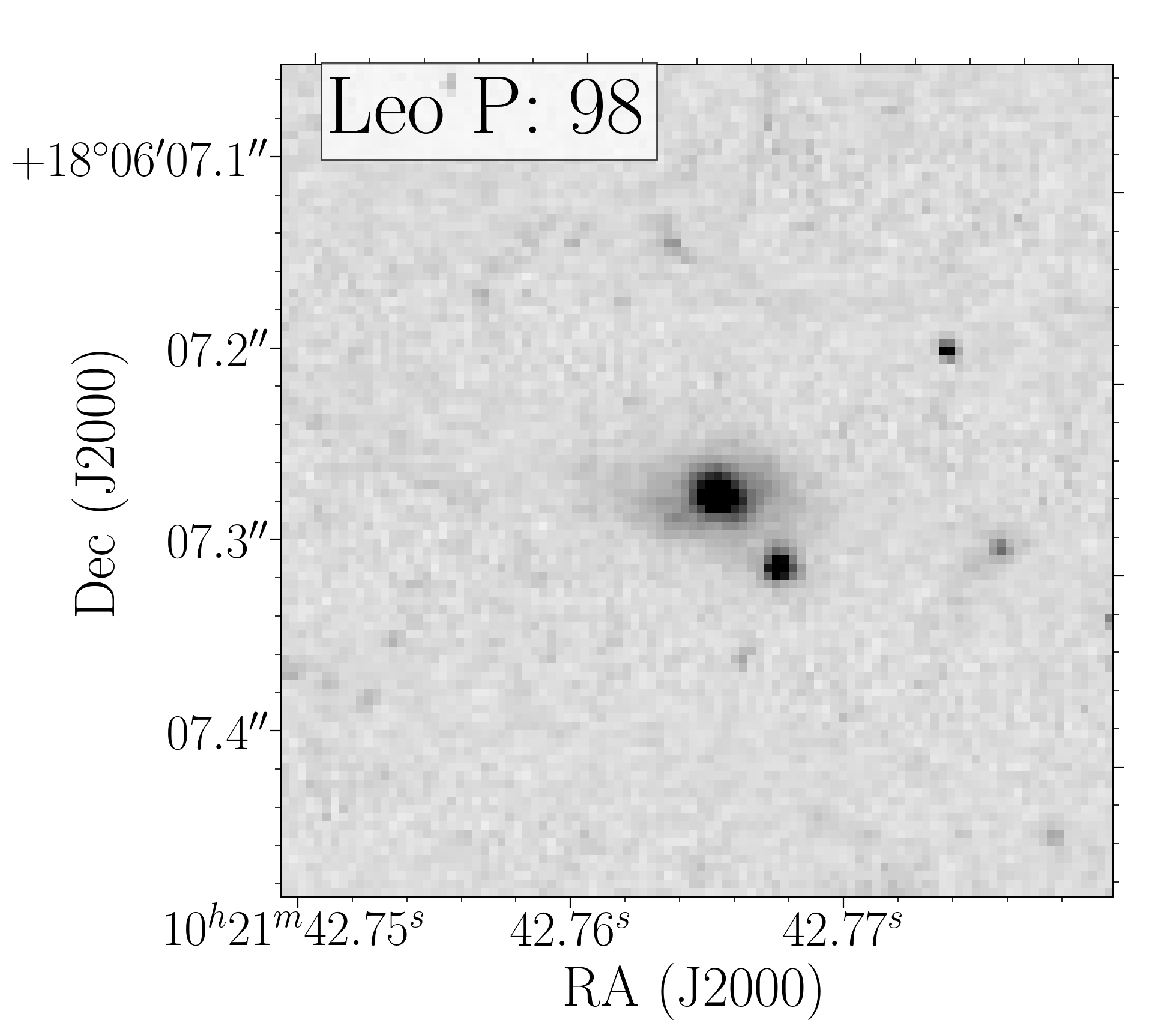} \\
 \includegraphics[width=0.32\linewidth]{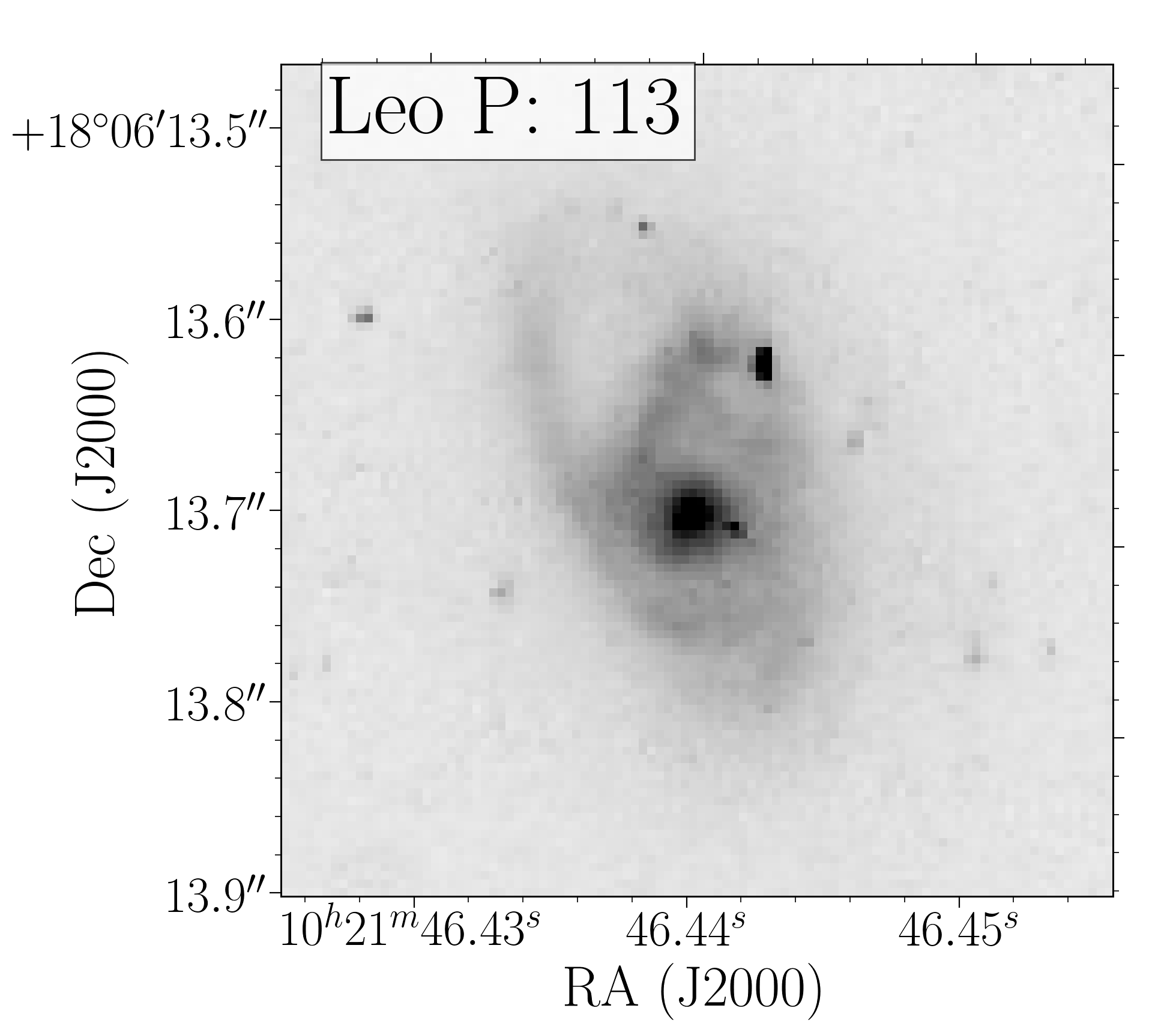} 
 \includegraphics[width=0.32\linewidth]{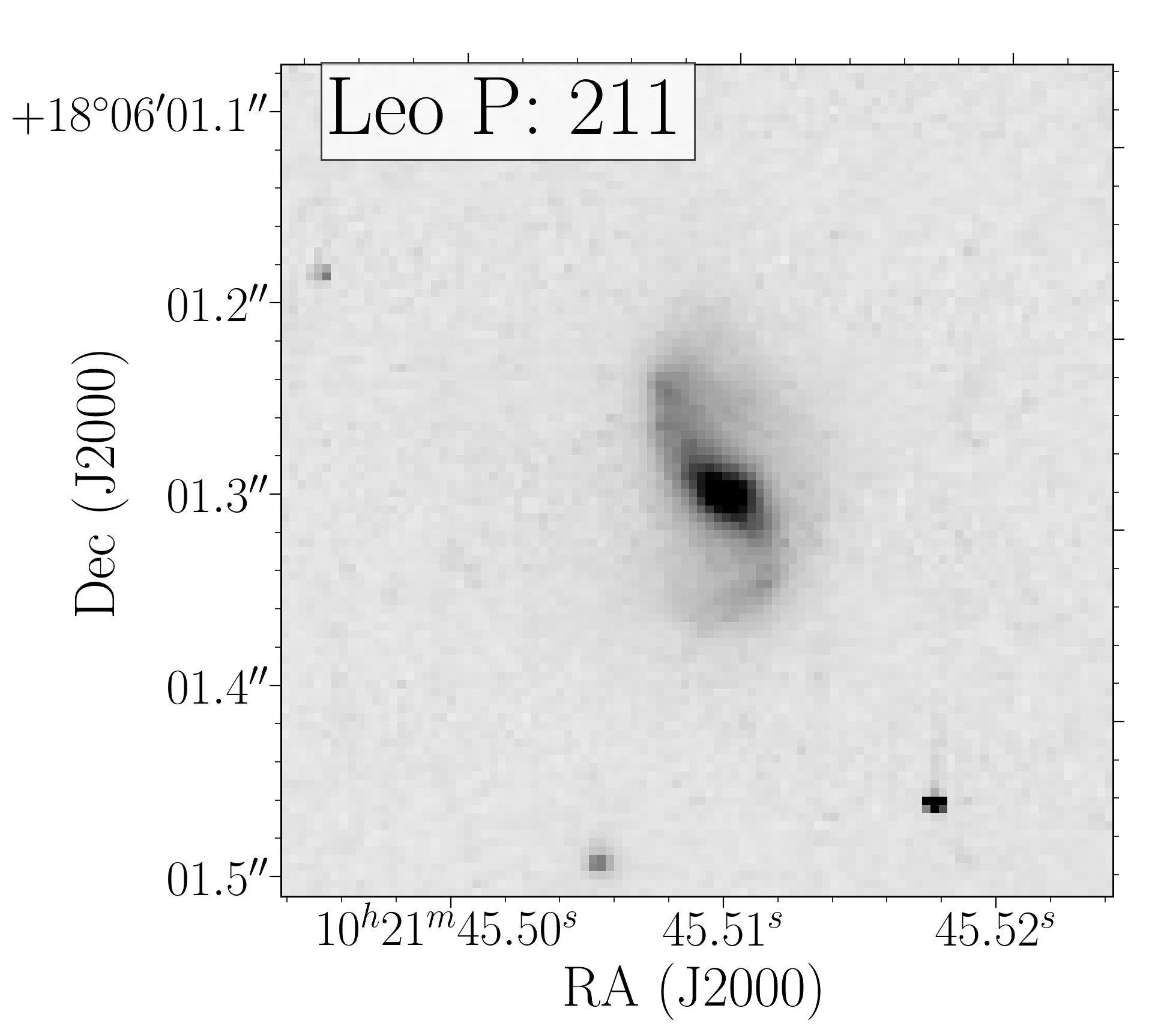}
 \includegraphics[width=0.32\linewidth]{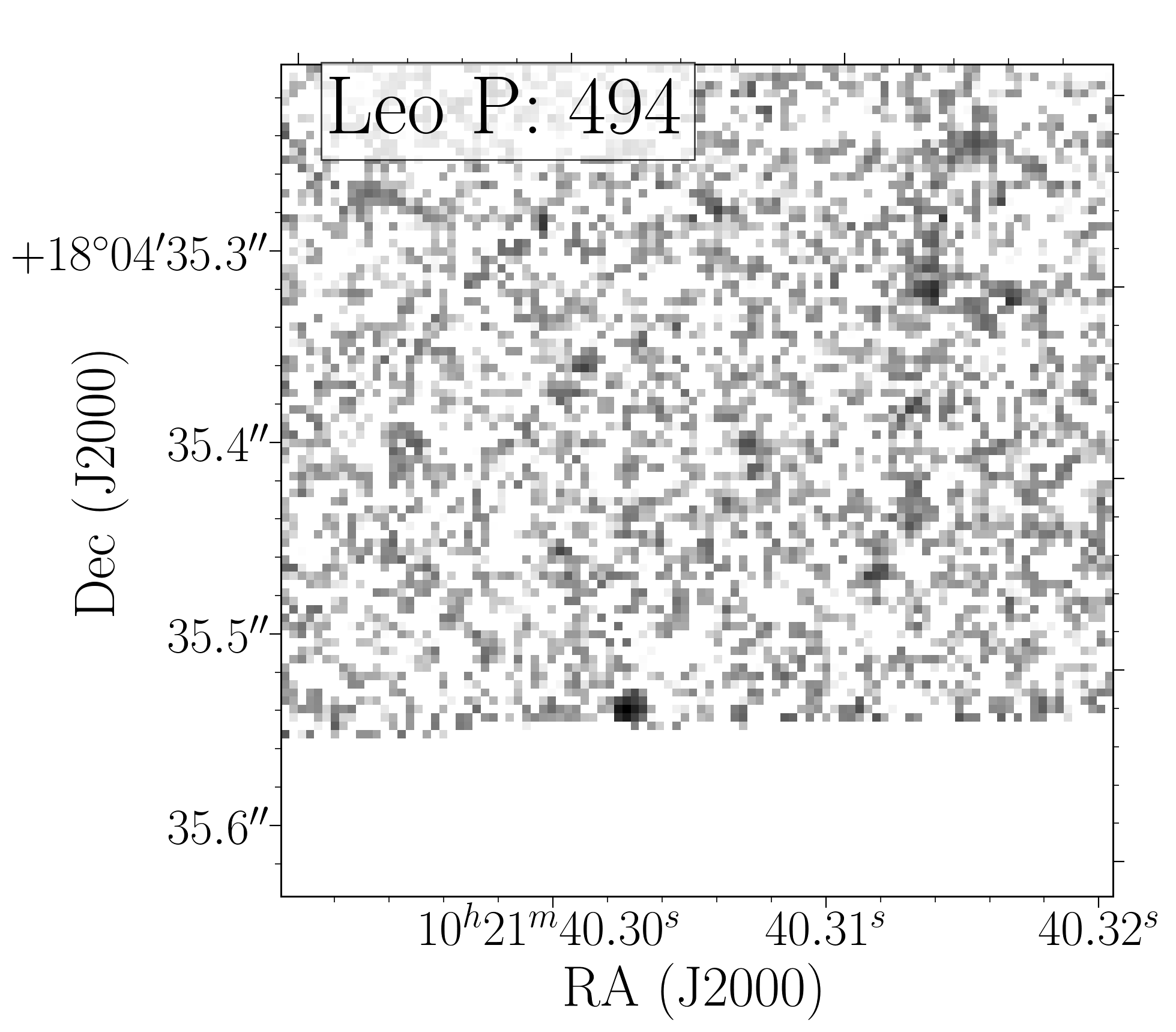} \\
 \includegraphics[width=0.32\linewidth]{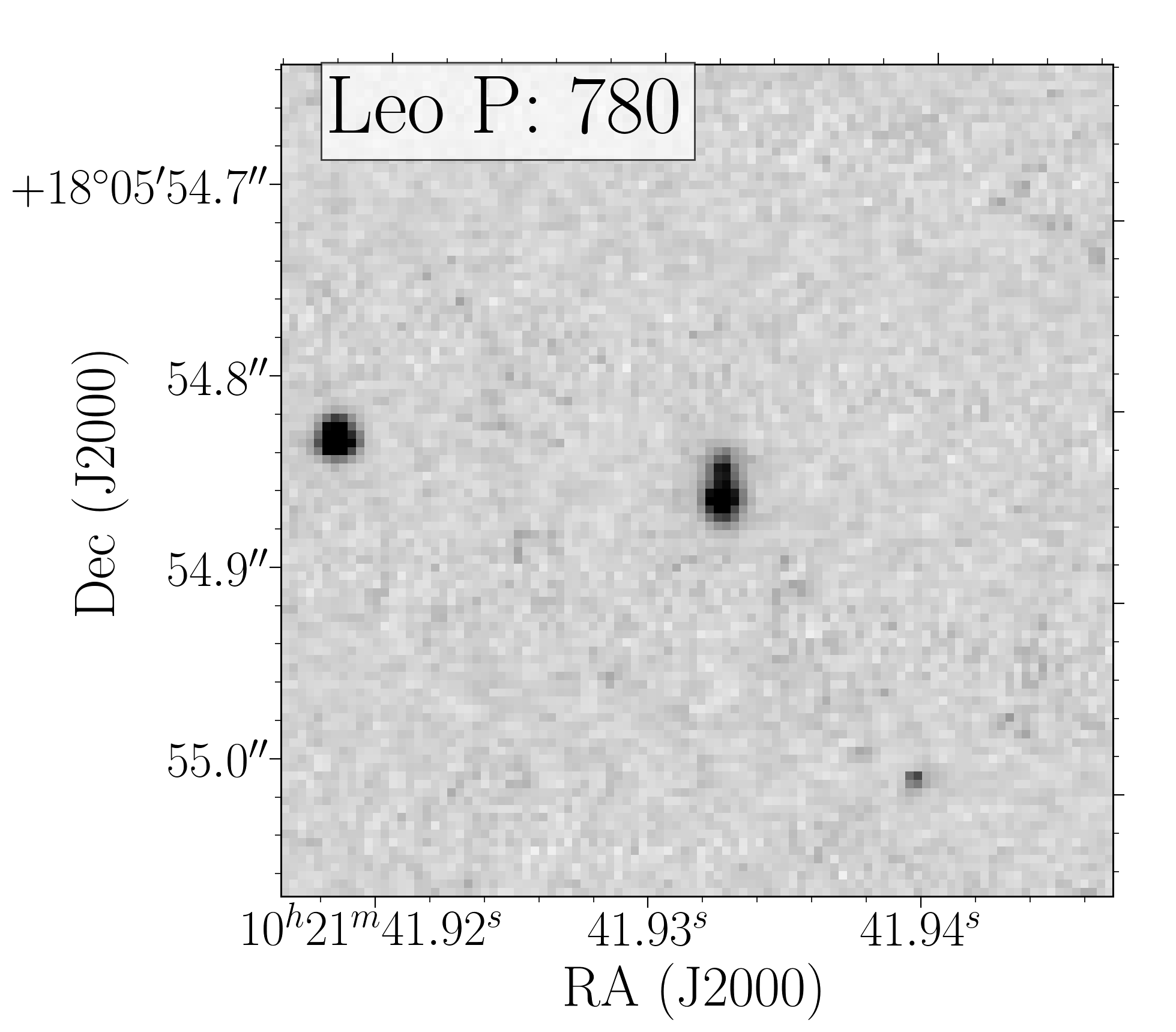} 
 \includegraphics[width=0.32\linewidth]{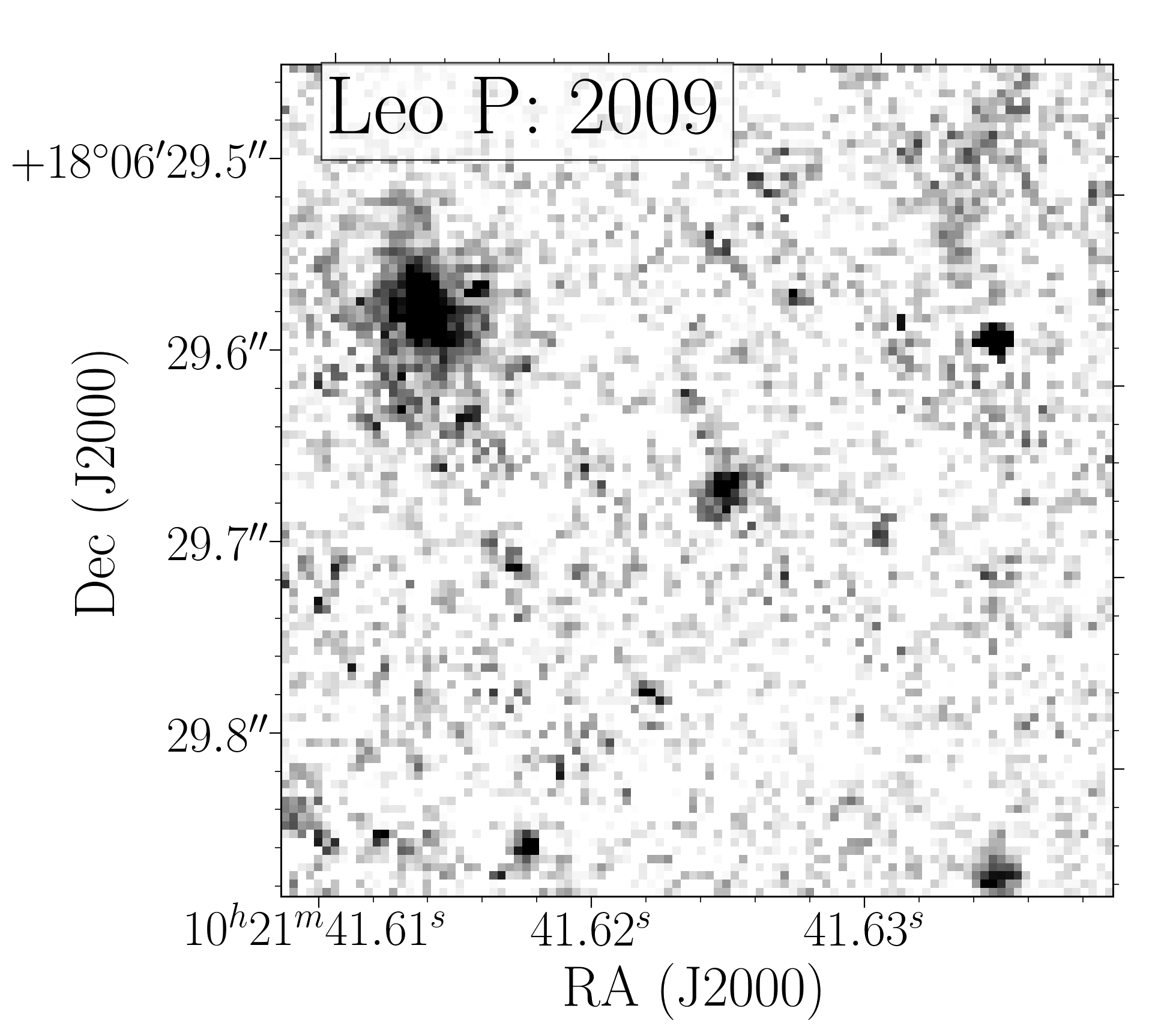}
 \includegraphics[width=0.32\linewidth]{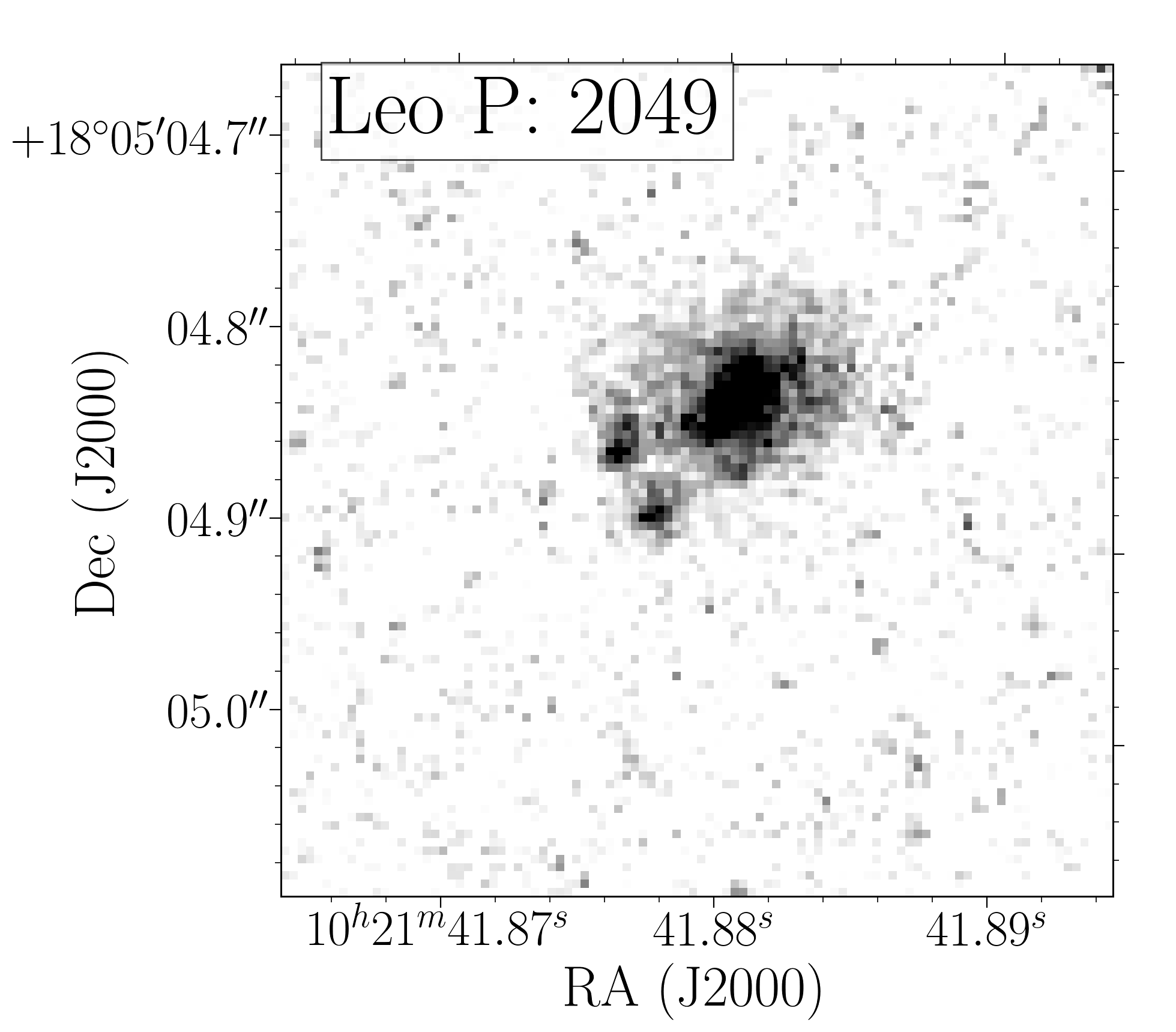} \\
 \caption{$HST/ACS$ F814W mosaic cutouts of sources above the TRGB or reddened, but that we have not classified as foreground sources or AGB candidates based on a visual inspection of the image. Source 494 is not likely to be a real source but rather the result of noise near the edge of the image. Source 2009 is the source at the center of the image.} 
 \label{fig: stamps} 
\end{figure*}

\begin{figure*}
 \includegraphics[width=0.32\linewidth]{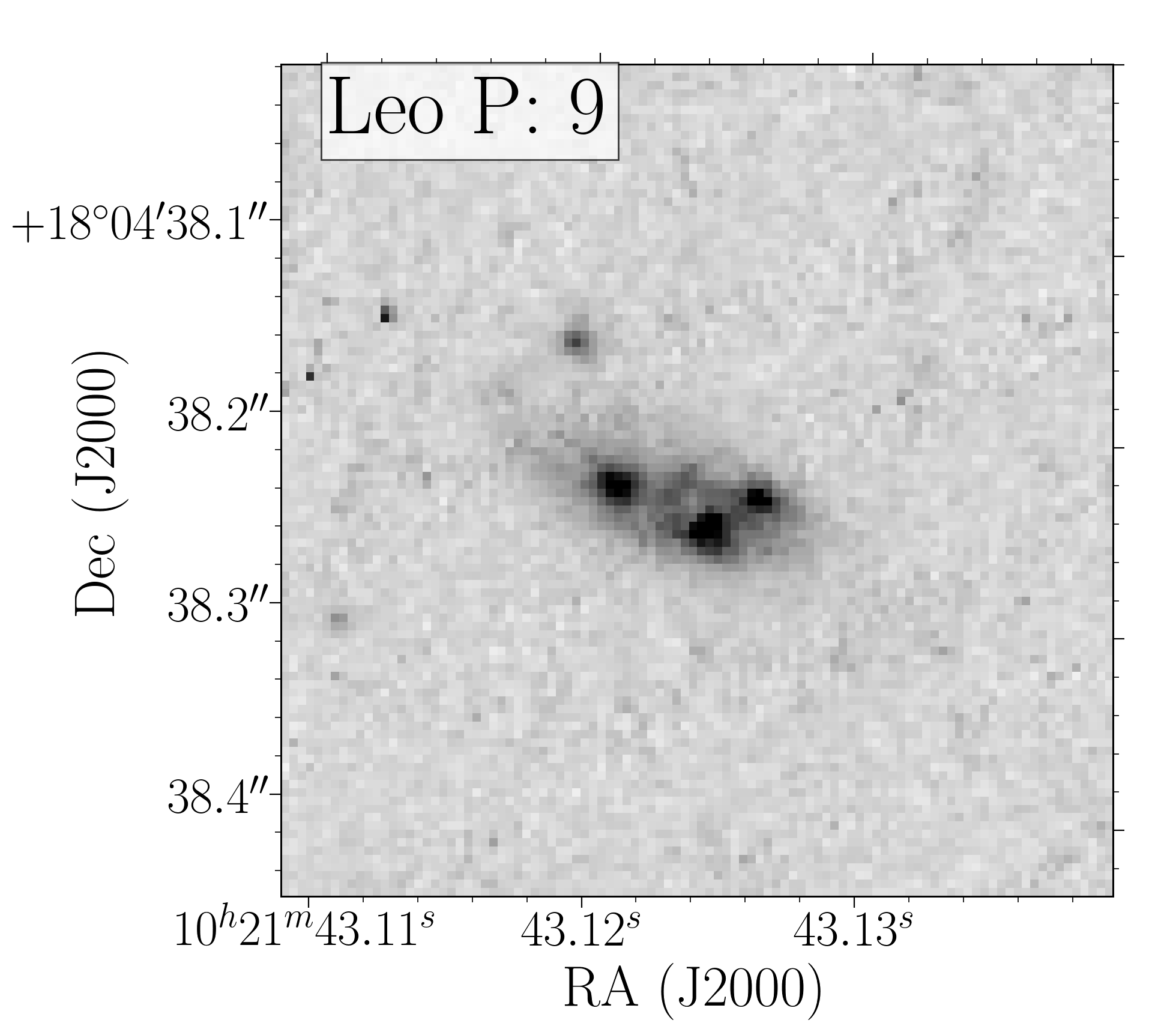}
 \includegraphics[width=0.32\linewidth]{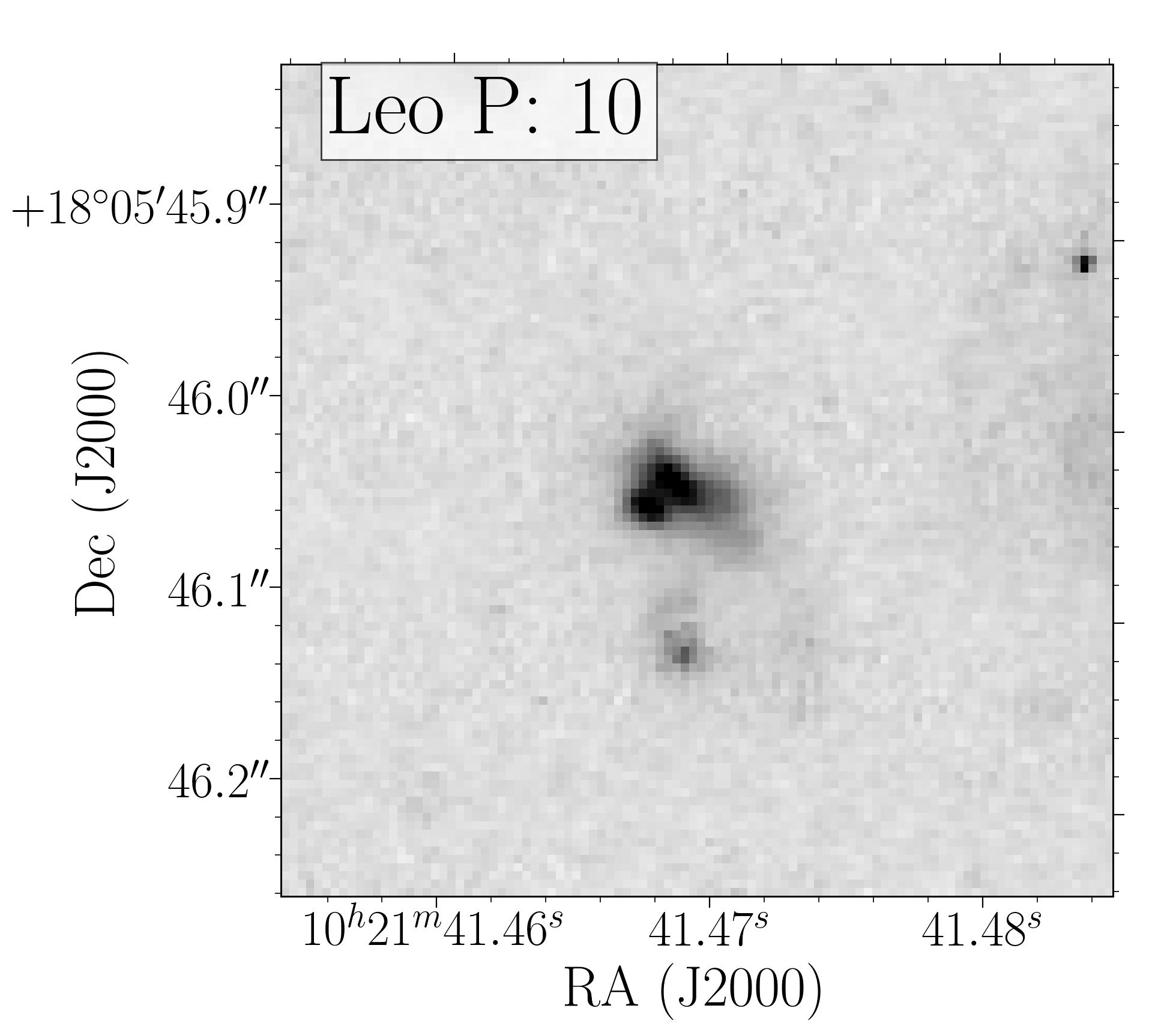} 
 \includegraphics[width=0.32\linewidth]{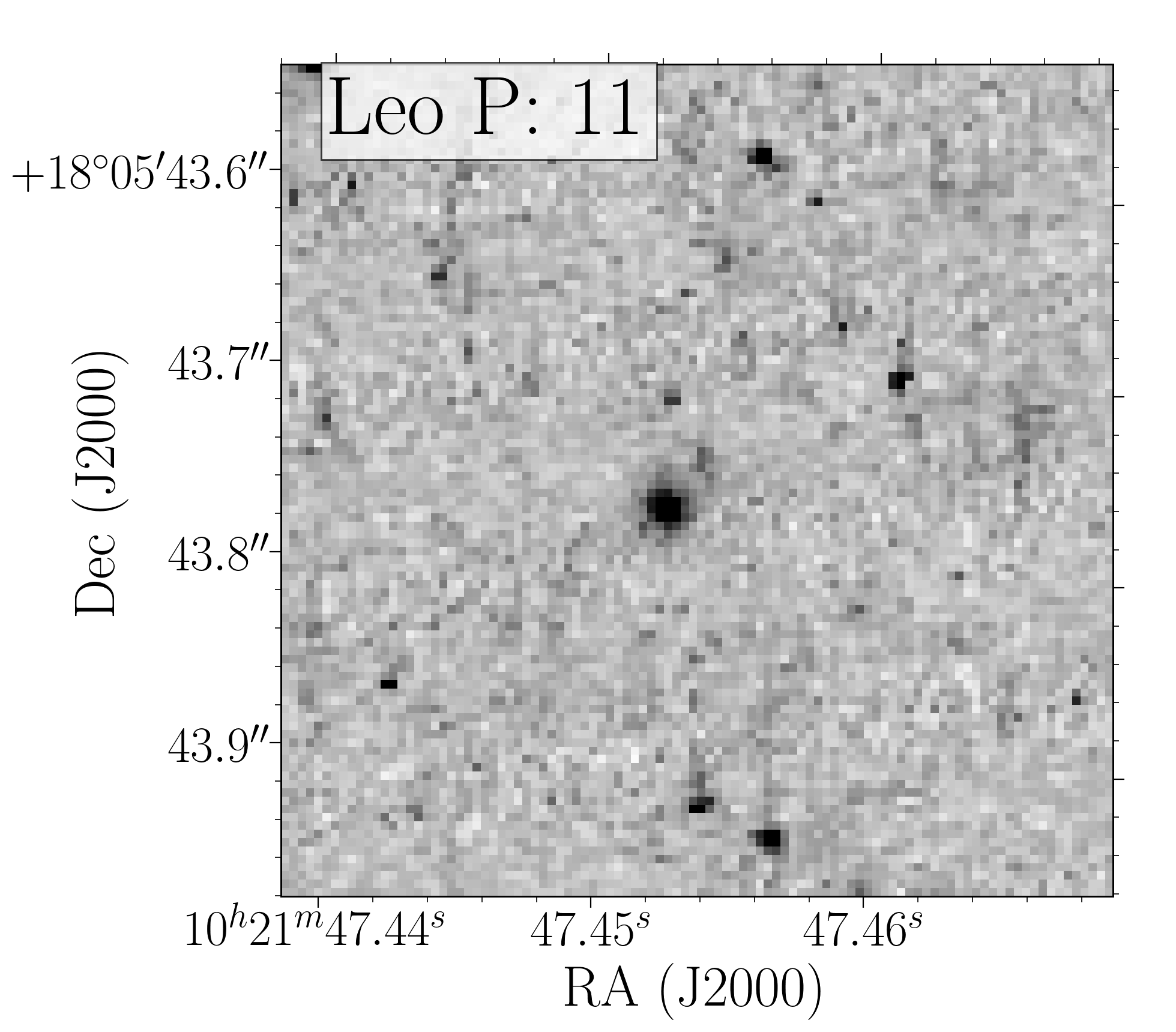} \\

 \includegraphics[width=0.32\linewidth]{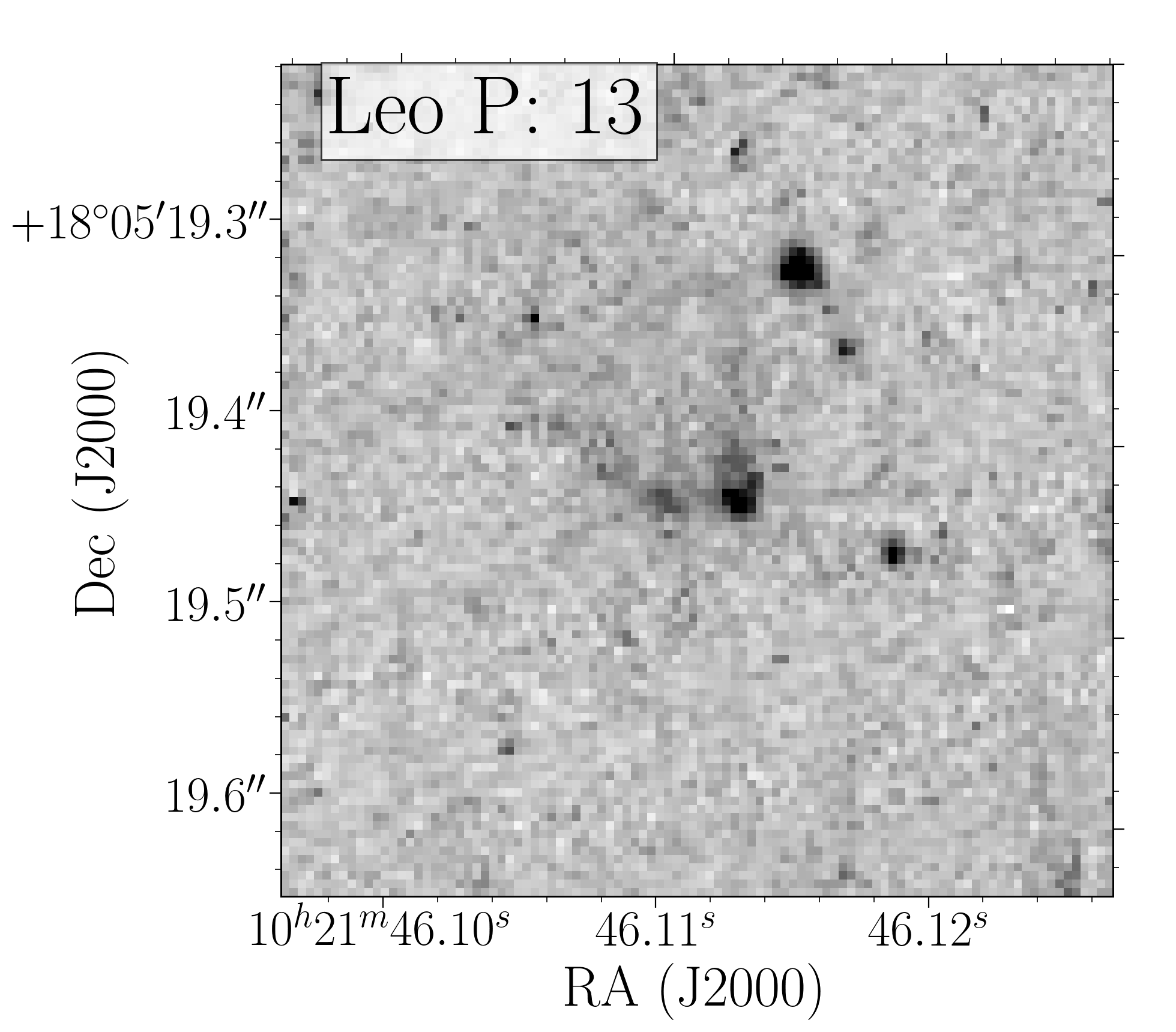}
 \includegraphics[width=0.32\linewidth]{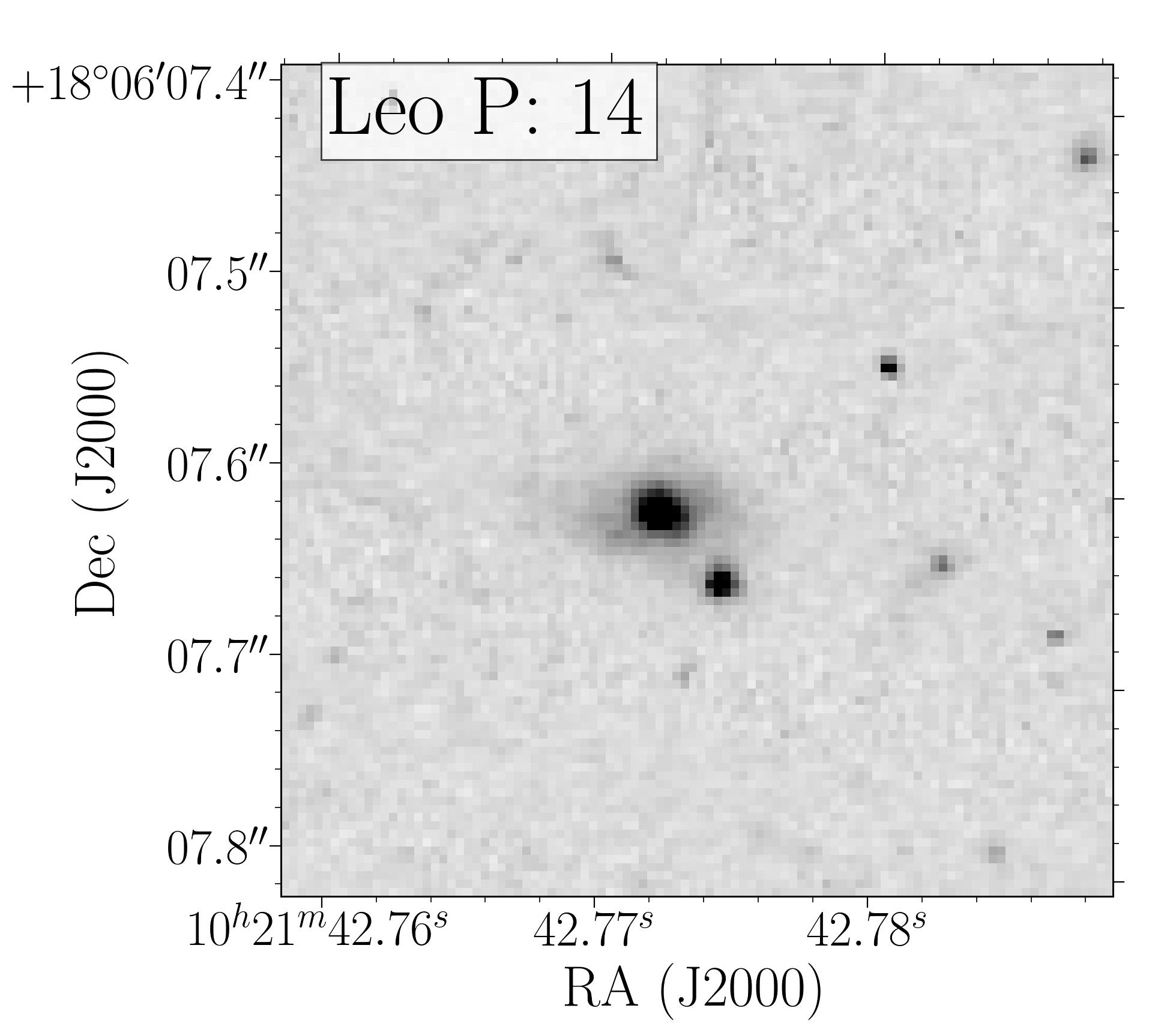} 
 \includegraphics[width=0.32\linewidth]{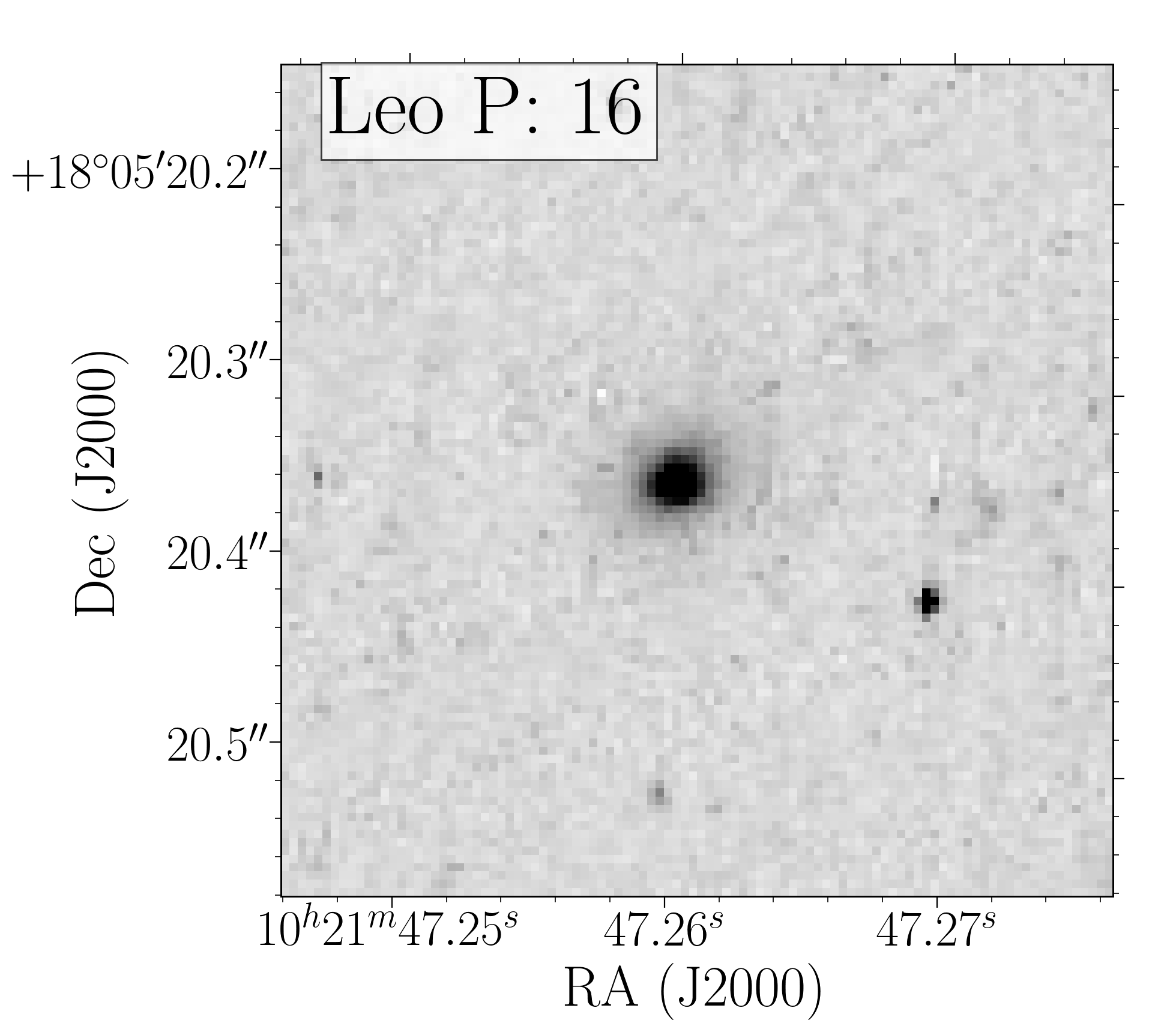} \\

 \includegraphics[width=0.32\linewidth]{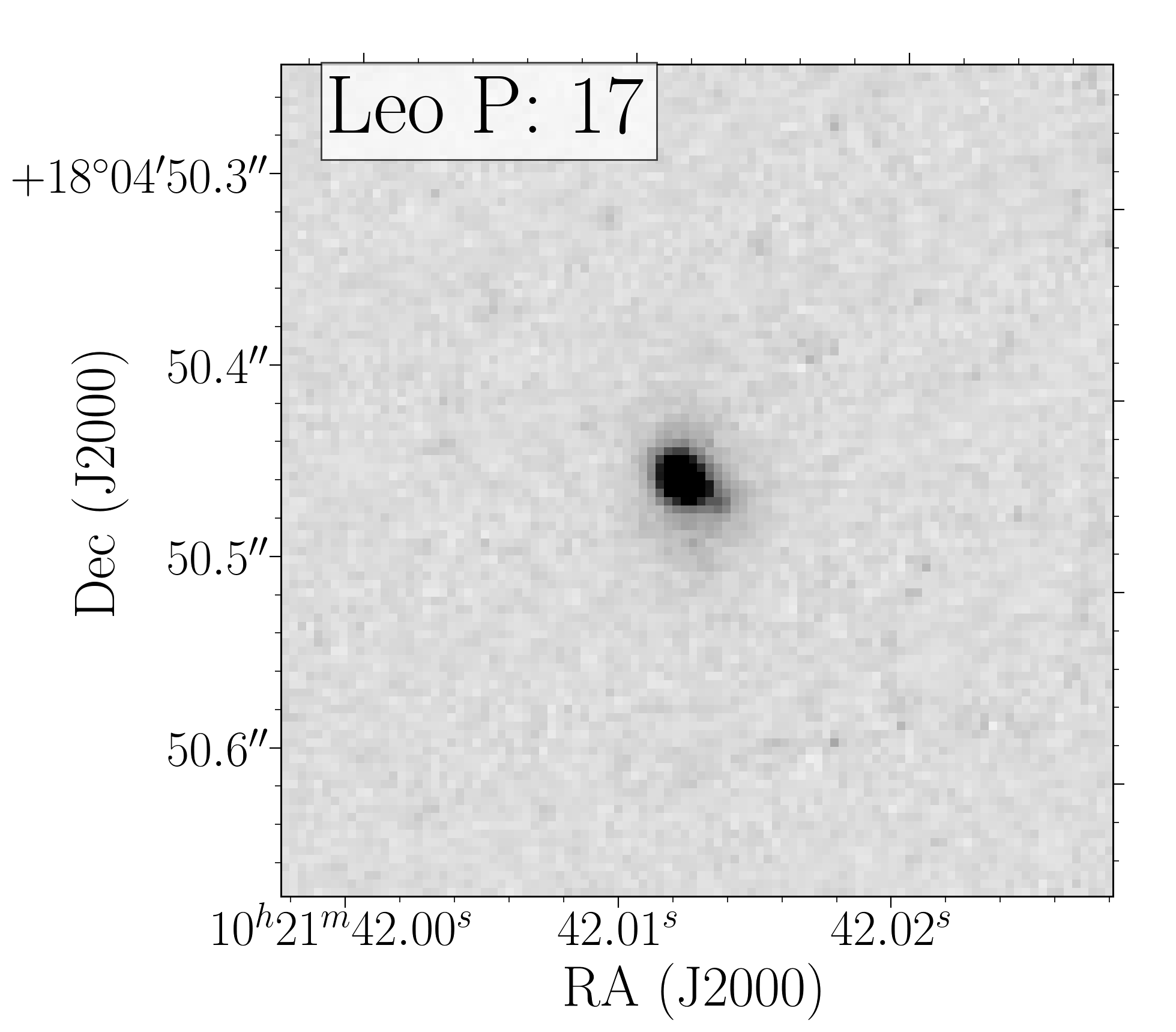}
 \includegraphics[width=0.32\linewidth]{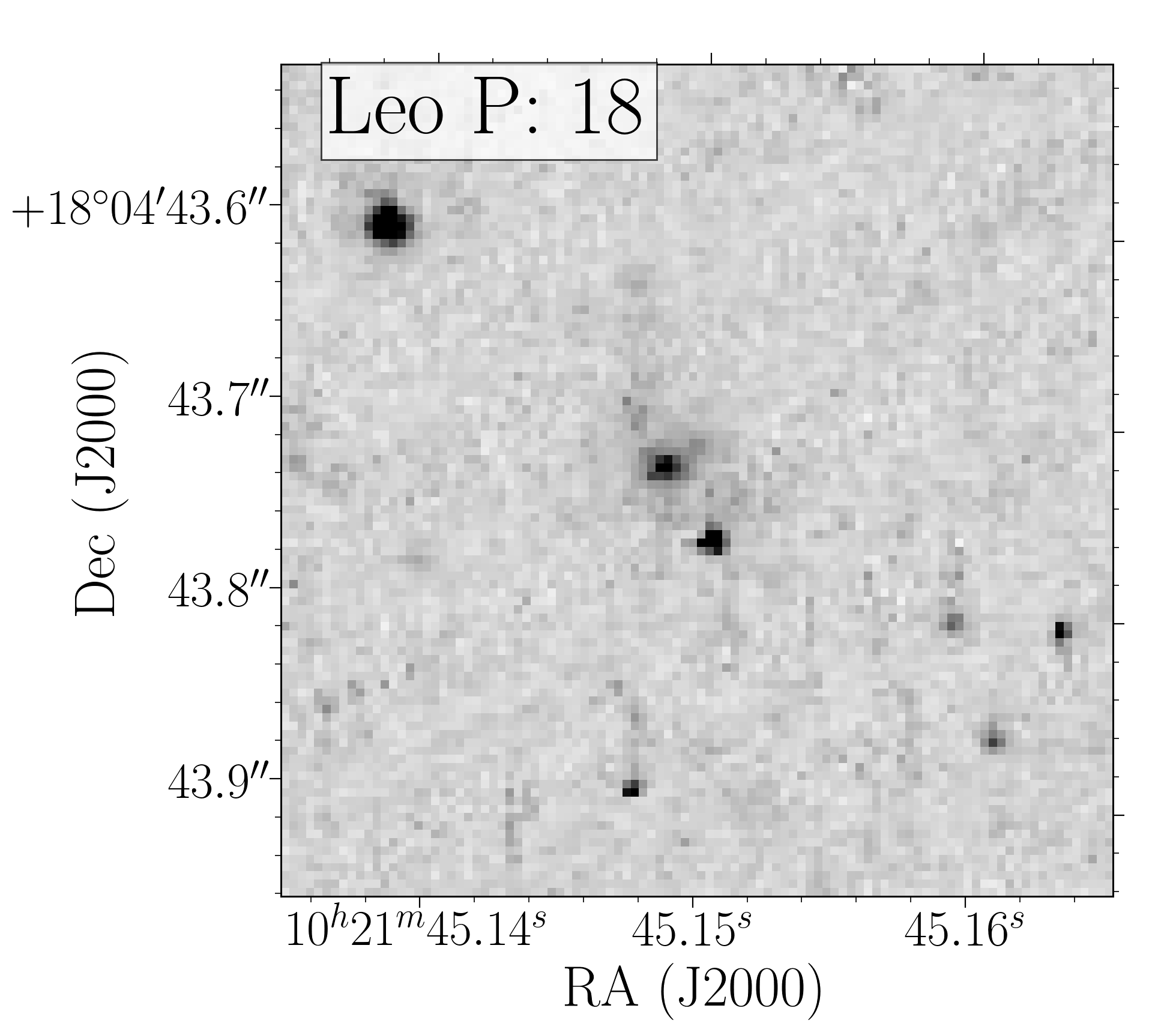} 
 \includegraphics[width=0.32\linewidth]{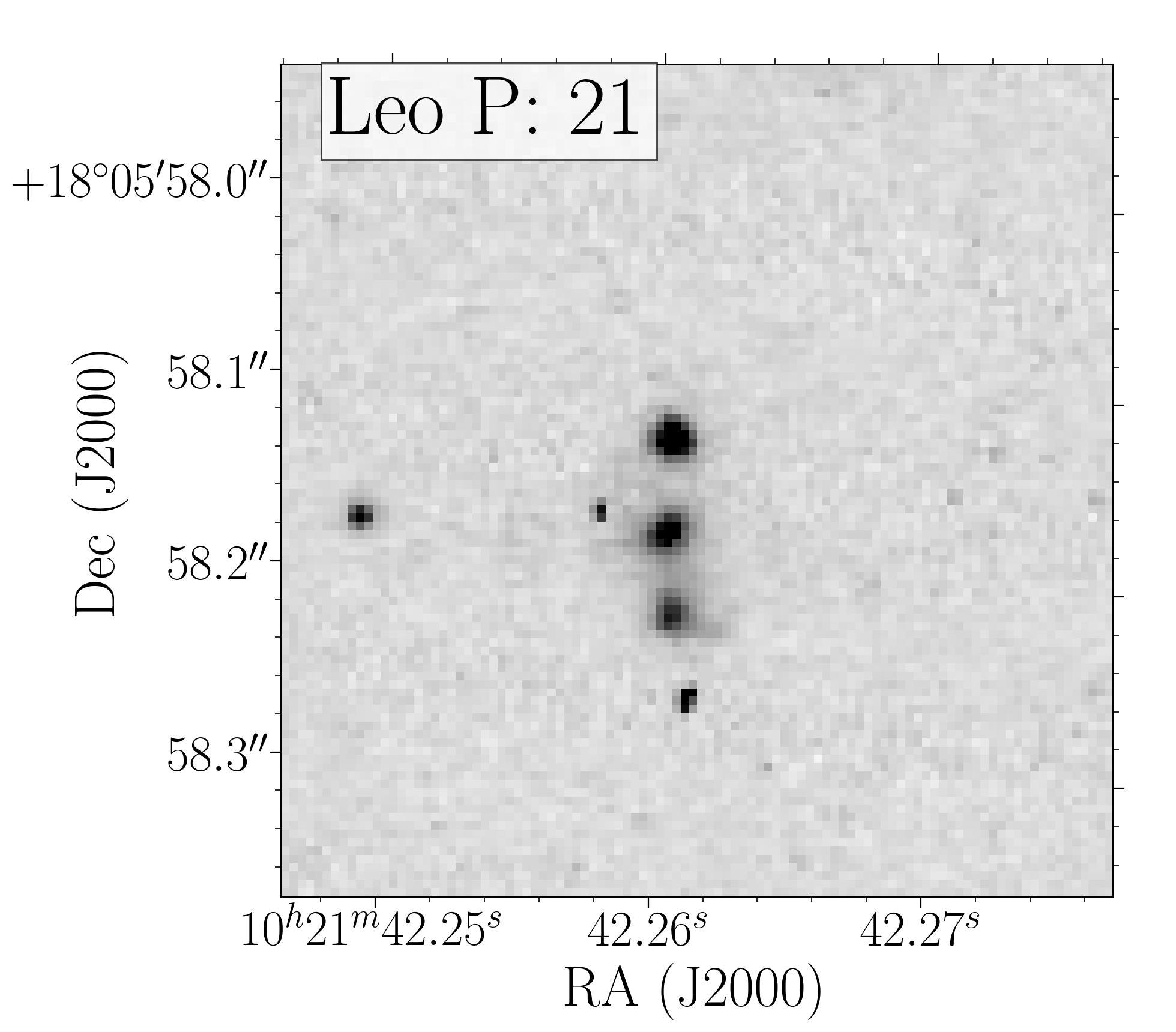} \\

 \caption{$HST/ACS$ F814W mosaic cutouts of sources that were determined AGB candidates by \citet{Lee2016} but that we have subsequently categorized as extended sources.}
 \label{fig: gal stamps} 
\end{figure*}

\begin{enumerate} 

\item Several sources that were initially classified as AGB stars in \citet{Lee2016} have now been reclassified as extended sources as a result of the higher-resolution imaging (Table \ref{table:photometry}; Figure \ref{fig: gal stamps}). 

\item AGB candidates that met out TRGB criteria and color, sharpness, and crowding cuts, but not our image quality criteria are shown in Table \ref{table:garbage_photometry} (extended objects are shown in Figure \ref{fig: stamps}). 

\item Table \ref{table:garbage_photometry} shows the photometry for sources that we have classified or reclassified as foreground stars or extended sources. 

\end{enumerate}

There are four sources that fall in the ``foreground'' region of our color-color diagram in Figure \ref{fig: trgb_knee} that are not listed as foreground stars in Table \ref{table:garbage_photometry}. Three of these are listed in Table \ref{table:photometry} as they were previously categorized as AGB candidates, and one was below the TRGB and has been found to be a galaxy (Source 113).

\begin{deluxetable*}{ccccccc}[b]
\tabletypesize{\small}
\tablecolumns{7}
\tablecaption{Leo P Additional Photometry \label{table:garbage_photometry}}

\tablehead{
\multirow{2}{*}{ID} &
\colhead{RA} & 
\colhead{Dec} & 
\colhead{F127M} &
\colhead{F139M} &
\colhead{F153M} &
\multirow{2}{*}{Type} \\
&
(J2000) &
(J2000) &
{\scriptsize (1.27$\mu$m)}&
{\scriptsize (1.39)}&
{\scriptsize (1.53)}&
}

\startdata
23 & 155.4360246 & 18.0828429 & 13.80 & 13.66 & 13.54 & K/FG \\
24 & 155.4305265 & 18.0842080 & 15.74 & 15.56 & 15.35 & K/FG \\
25 & 155.4256185 & 18.0944481 & 16.04 & 15.89 & 15.69 & K/FG \\
26 & 155.4302984 & 18.0870885 & 16.15 & 16.08 & 15.90 & K/FG \\
27 & 155.4256132 & 18.0943271 & 16.87 & 16.74 & 16.56 & K/FG \\
28 & 155.4312416 & 18.0860352 & 18.57 & 18.46 & 18.28 & K/FG \\
29 & 155.4558705 & 18.0853354 & 19.76 & 19.60 & 19.41 & K/FG \\
31 & 155.4268835 & 18.0764639 & 21.35 & 21.22 & 21.16 & K/FG \\
45 & 155.4296613 & 18.0882035 & 20.55 & 20.23 & 19.95 & \ldots \\
52 & 155.4450924 & 18.0902926 & 20.71 & 20.56 & 20.13 & Ext \\
98 & 155.4289005 & 18.1027023 & 21.40 & 20.92 & 20.77 & Ext \\
211 & 155.4403415 & 18.1010466 & 21.78 & 21.43 & 21.15 & Ext \\
494 & 155.4186578 & 18.0772001 & 22.72 & 22.68 & 21.36 & \ldots \\
780 & 155.4254298 & 18.0992646 & 23.94 & 23.83 & 21.67 & \ldots \\
2009 & 155.4241478 & 18.1089321 & 23.36 & 23.12 & 22.25 & \ldots \\
2049 & 155.4252101 & 18.0853781 & 24.34 & 22.48 & 22.99 & Ext 
\enddata 

\end{deluxetable*}

\end{document}